\RequirePackage{amsmath}
\documentclass[epj, nopacs]{svjour}

\usepackage[utf8]{inputenc}

\usepackage[export]{adjustbox}
\usepackage{graphicx}
\usepackage{times}

\usepackage{csquotes}
\usepackage[english]{babel}

\DeclareMathSymbol{\varOmega}{\mathalpha}{operators}{"0A}
\usepackage{siunitx}
\usepackage{xspace}
\usepackage[version=4]{mhchem}
\usepackage{physics}
\usepackage{xfrac}
\usepackage{amsbsy}
\usepackage[artemisia]{textgreek}
\usepackage{upgreek}
\usepackage{amssymb}
\usepackage{subcaption}
\usepackage{xcolor}
\usepackage{comment}

\usepackage{diagbox}

\usepackage{wrapfig}
\usepackage{refcheck} 

\newcommand{\eq}{\begin{equation}}
\newcommand{\eeq}{\end{equation}}

\newcommand{\Kaon}{K\xspace}

\newcommand{\prot}{p\xspace}

\newcommand{\pim}{\textpi$^{-}$\xspace}
\newcommand{\pip}{\textpi$^{+}$\xspace}

\newcommand{\Kp}{\Kaon{}$^{+}$\xspace}

\newcommand{\Szero}{\textSigma$^0$\xspace}

\newcommand{\Sstar}{\textSigma(1385)\xspace}

\newcommand{\Lzero}{\textLambda\xspace}
\newcommand{\Lstar}{\textLambda(1405)\xspace}
\newcommand{\Lstard}{\textLambda(1520)\xspace}

\newcommand{\pdecay}{$\mathrm{p_{decay}}$\xspace}
\newcommand{\SSignal}{$\mathrm{pp \rightarrow pK^{+}\Sigma^{0}}$\xspace}
\newcommand{\LSignal}{$\mathrm{pp \to pK^{+}\Lambda}$\xspace}
\newcommand{\LBack}{$\mathrm{pp \to pK^{+}\Lambda\pi^{0}}$\xspace}
\newcommand{\GeV}{$\mathrm{GeV/c^{2}}$\xspace}
\newcommand{\spin}{$\mathrm{J^{P}}$}

\DeclareSIUnit\clight{\text{\ensuremath{c}}}

\DeclareSIUnit\mub{\micro\barn}

\DeclareSIUnit\gev{\giga\electronvolt}
\DeclareSIUnit\gevc{\giga\electronvolt\per\clight}
\DeclareSIUnit\gevsc{\giga\electronvolt\per\clight\squared}
\DeclareSIUnit\sgevsc{\giga\electronvolt\squared\per\clight\tothe{4}}

\DeclareSIUnit\mev{\mega\electronvolt}
\DeclareSIUnit\mevc{\mega\electronvolt\per\clight}
\DeclareSIUnit\mevsc{\mega\electronvolt\per\clight\squared}
\DeclareSIUnit\smevsc{\mega\electronvolt\squared\per\clight\tothe{4}}

\DeclareSIUnit\kev{\kilo\electronvolt}
\DeclareSIUnit\kevc{\kilo\electronvolt\per\clight}
\DeclareSIUnit\kevsc{\kilo\electronvolt\per\clight\squared}

\usepackage[inline]{enumitem}

\usepackage{tabularx}
\usepackage{booktabs}
\usepackage{multicol}

\newcolumntype{L}[1]{>{\hsize=#1\hsize\raggedright\arraybackslash}X}%
\newcolumntype{R}[1]{>{\hsize=#1\hsize\raggedleft\arraybackslash}X}%
\newcolumntype{C}[1]{>{\hsize=#1\hsize\centering\arraybackslash}X}%

\usepackage{hyperref}
\hypersetup{
  colorlinks=true,
  linkcolor=red,     
  citecolor=blue,    
  filecolor=magenta, 
  urlcolor=cyan,     
}

\usepackage[
  backend=biber,
  bibstyle=numeric-epj,
  alldates=iso,
  seconds=true,
  doi=true,
  url=false,
  isbn=false,
  eprint=true,
  giveninits=true,
]{biblatex}

\DeclareSourcemap{
 \maps[datatype=bibtex,overwrite=true]{
  \map{
    \step[fieldsource=Collaboration, final=true]
    \step[fieldset=usera, origfieldval, final=true]
  }
 }
}

\DeclareFieldFormat{pages}{#1}
\renewbibmacro{in:}{}
\renewbibmacro*{doi+eprint+url}{%
    \printfield{doi}%
    \newunit\newblock%
    \iffieldundef{doi}{%
    \iftoggle{bbx:eprint}{%
        \usebibmacro{eprint}%
    }{}%
    }{}%
    \newunit\newblock%
    \iffieldundef{doi}{%
        \usebibmacro{url+urldate}}%
        {}%
    }

\bibliography{biblio}

\makeatletter
\def\cl@chapter{\@elt {theorem}}
\makeatother

\usepackage[capitalize]{cleveref}
\usepackage[switch]{lineno}

\begin{document}
\title{Investigation of the $\mathbf{\Sigma^{0}}$ Production Mechanism in p(3.5 GeV)+p Collisions}

\author{
R.~Abou Yassine$^{6,13}$ \and
O.~Arnold$^{10,9}$ \and
M.~Becker$^{11}$ \and
P.~Bergmann$^{5}$ \and
A.~Blanco$^{1}$ \and
C.~Blum$^{8}$ \and
M.~B\"{o}hmer$^{10}$ \and
N.~Carolino$^{1}$ \and
L.~Chlad$^{14,c}$ \and
P.~Chudoba$^{14}$ \and
I.~Ciepa{\l}$^{3}$ \and
J.~Dreyer$^{7}$ \and
W.~Esmail$^{5}$ \and
L.~Fabbietti$^{10,9}$ \and
P.~Fonte$^{1,a}$ \and
J.~Friese$^{10}$ \and
I.~Fr\"{o}hlich$^{8}$ \and
T.~Galatyuk$^{6,5}$ \and
J.~A.~Garz\'{o}n$^{15}$ \and
M.~Grunwald$^{17}$ \and
M.~Gumberidze$^{5}$ \and
S.~Harabasz$^{6,b}$ \and
C.~H\"{o}hne$^{11,5}$ \and
F.~Hojeij$^{13}$ \and
R.~Holzmann$^{5}$ \and
H.~Huck$^{8}$ \and
M.~Idzik$^{2}$ \and
B.~K\"{a}mpfer$^{7,c}$ \and
B.~Kardan$^{8}$ \and
V.~Kedych$^{6}$ \and
I.~Koenig$^{5}$ \and
W.~Koenig$^{5}$ \and
M.~Kohls$^{8}$ \and
J.~Kolas$^{17}$ \and
G.~Korcyl$^{4}$ \and
G.~Kornakov$^{17}$ \and
R.~Kotte$^{7}$ \and
W.~Krueger$^{6}$ \and
A.~Kugler$^{14}$ \and
T.~Kunz$^{10}$ \and
R.~Lalik$^{4}$ \and
F.~Linz$^{6,5}$ \and
L.~Lopes$^{1}$ \and
M.~Lorenz$^{8}$ \and
A.~Malige$^{4}$ \and
J.~Markert$^{5}$ \and
V.~Metag$^{11}$ \and
J.~Michel$^{8}$ \and
A. Molenda$^{2}$ \and
C.~M\"{u}ntz$^{8}$ \and
M.~Nabroth$^{8}$ \and
L.~Naumann$^{7}$ \and
K.~Nowakowski$^{4}$ \and
J.~Orliński$^{16}$ \and
J.-H.~Otto$^{11}$ \and
Y.~Parpottas$^{12}$ \and
M.~Parschau$^{8}$ \and
V.~Pechenov$^{5}$ \and
O.~Pechenova$^{5}$ \and
K.~Piasecki$^{16}$ \and
J.~Pietraszko$^{5}$ \and
A.~Prozorov$^{14,d}$ \and
W.~Przygoda$^{4}$ \and
B.~Ramstein$^{13}$ \and
N.~Rathod$^{17}$ \and
J.~Ritman$^{5}$ \and
A.~Rost$^{6,5}$ \and
A.~Rustamov$^{5}$ \and
P.~Salabura$^{4}$ \and
N.~Schild$^{6}$ \and
E.~Schwab$^{5}$ \and
F.~Seck$^{6}$ \and
U.~Singh$^{4}$ \and
S.~Spies$^{8}$ \and
M.~Stefaniak$^{17,5}$ \and
H.~Str\"{o}bele$^{8}$ \and
J.~Stroth$^{8,5}$ \and
C.~Sturm$^{5}$ \and
K.~Sumara$^{4}$ \and
O.~Svoboda$^{14}$ \and
M.~Szala$^{8}$ \and
P.~Tlusty$^{14}$ \and
M.~Traxler$^{5}$ \and
H.~Tsertos$^{12}$ \and
V.~Wagner$^{14}$ \and
A.A.~Weber$^{11}$ \and
C.~Wendisch$^{5}$ \and
H.P.~Zbroszczyk$^{17}$ \and
E.~Zherebtsova$^{5,e}$ \and
M.~Zielinski$^{4}$ \and
P.~Zumbruch$^{5}$ 
(HADES collaboration)
}

\institute{
$^{1}$ LIP-Laborat\'{o}rio de Instrumenta\c{c}\~{a}o e F\'{\i}sica Experimental de Part\'{\i}culas 3004-516 Coimbra, Portugal \\
$^{2}$ AGH University of Science and Technology, Faculty of Physics and Applied Computer Science, 30-059 Krak\'{o}w, Poland \\
$^{3}$ Institute of Nuclear Physics, Polish Academy of Sciences, 31342~Krak\'{o}w, Poland \\
$^{4}$ Smoluchowski Institute of Physics, Jagiellonian University of Cracow, 30-059 Krak\'{o}w, Poland \\
$^{5}$ GSI Helmholtzzentrum f\"{u}r Schwerionenforschung GmbH, 64291 Darmstadt, Germany \\
$^{6}$ Technische Universit\"{a}t Darmstadt, 64289 Darmstadt, Germany \\
$^{7}$ Institut f\"{u}r Strahlenphysik, Helmholtz-Zentrum Dresden-Rossendorf, 01314 Dresden, Germany \\
$^{8}$ Institut f\"{u}r Kernphysik, Goethe-Universit\"{a}t, 60438 Frankfurt, Germany \\
$^{9}$ Excellence Cluster ’Origin and Structure of the Universe’, 85748 Garching, Germany \\
$^{10}$ Physik Department E62, Technische Universit\"{a}t M\"{u}nchen, 85748 Garching, Germany \\
$^{11}$ II.Physikalisches Institut, Justus Liebig Universit\"{a}t Giessen, 35392 Giessen, Germany \\
$^{12}$ Department of Physics, University of Cyprus, 1678 Nicosia, Cyprus \\
$^{13}$ Laboratoire de Physique des 2 infinis Irène Joliot-Curie, Université Paris-Saclay, CNRS-IN2P3., F-91405 Orsay, France \\
$^{14}$ Nuclear Physics Institute, The Czech Academy of Sciences, 25068 Rez, Czech Republic \\
$^{15}$ LabCAF. F. Física, Univ. de Santiago de Compostela, 15706 Santiago de Compostela, Spain \\
$^{16}$ Uniwersytet Warszawski - Instytut Fizyki Do\'{s}wiadczalnej, 02-093 Warszawa, Poland \\
$^{17}$ Warsaw University of Technology, 00-662 Warsaw, Poland \\
\\
$^{a}$ also at Coimbra Polytechnic - ISEC, Coimbra, Portugal \\
$^{b}$ also at Helmholtz Research Academy Hesse for FAIR (HFHF), Campus Darmstadt, 64390 Darmstadt, Germany \\
$^{c}$ also at Technische Universit\"{a}t Dresden, 01062 Dresden, Germany \\
$^{d}$ also at Charles University, Faculty of Mathematics and Physics, 12116 Prague, Czech Republic \\
$^{e}$ also at University of Wrocław, 50-204 Wrocław, Poland \\
\\
\email{hades-info@gsi.de}
}

\authorrunning{R.~Abou Yassine et al.} 
\titlerunning{Investigation of the \Szero Production Mechanism in p(3.5 GeV)+p Collisions}

\date{}

\abstract{
The production of \Szero hyperons in proton proton collisions at a beam kinetic energy of 3.5 GeV impinging on a liquid hydrogen target was investigated using data collected with the HADES setup. The total production cross section is found to be $\mathrm{\sigma (pK^{+}\Sigma^{0}) [\mu b] = 17.7 \pm 1.7 (stat) \pm 1.6 (syst)}$. Differential cross section distributions of the exclusive channel \SSignal were analyzed in the center-of-mass, Gottfried-Jackson and helicity reference frames for the first time at the excess energy of 556 MeV. The data support the interplay between pion and kaon exchange mechanisms and clearly demonstrate the contribution of interfering nucleon resonances decaying to $\mathrm{K^{+}\Sigma^{0}}$. The Bonn-Gatchina partial wave analysis was employed to analyse the data. Due to the limited statistics, it was not possible to obtain an unambiguous determination of the relative contribution of intermediate nucleon resonances to the final state. However nucleon resonances with masses around 1.710 \GeV ($\mathrm{N^{*}(1710)}$) and 1.900 \GeV ($\mathrm{N^{*}(1900)}$ or $\mathrm{\Delta^{*}(1900)}$) are preferred by the fit.
\PACS{
 {PACS-key}{describing text of that key} \and
 {PACS-key}{describing text of that key}
}
}
\maketitle


\section{Introduction}\label{sec:intro}
Strangeness production at intermediate energies in p+p and p+A collisions is of particular importance to the field of hadron physics. The production of baryons with strange quark content, i.e. hyperons, requires creating a new quark flavor, which can occur out of the vacuum from the quark sea in the colliding protons. The s-quark with mass ~ O($\Lambda_{QCD}$) is distinguished from the light (u, d) quark flavors but much smaller than heavy (c, t, b) flavors. The resulting (approximate) SU(3) flavor symmetry in the u-d-s sector is therefore still a cornerstone of hadron physics. Since the entrance channel in p+p and p+A collisions carries no net strangeness, the emergence of an s-quark can unravel much of the flavor dynamics in hadronic reactions. The flavor-conserving strong interaction process requires associate strangeness production, e.g. realized by simultaneous creation of a single-strange hyperon, such as \Lzero or \Szero and an associated kaon. Therefore, understanding the production mechanism of strange baryons near threshold deepens our knowledge of their internal structure and of the strong interaction in the non-perturbative regime. Strangeness production is also used as a probe to study hot and dense nuclear matter in heavy ion collisions both at medium-energy and in the late stages prior to freeze-out in high-energy collisions, e.g. at LHC \cite{Brown:1991ig}. 

The production of the \Lzero hyperon in p+p and p+A reactions near threshold has been studied extensively by many experiments including HADES \cite{AbdelBary:2010pc, Adamczewski-Musch:2016vrc, Agakishiev:2014kdy, Balewski:1991ns, Munzer:2017hbl}, yet there are only few experimental investigations on the \Szero hyperon \cite{AbdelBary:2010pc, Adamczewski-Musch:2017gjr}. Despite there are considerable experimental results and numerous dedicated theoretical investigations, the strangeness production mechanism is not yet well understood. In the context of the boson exchange model \cite{Chew:1958wd, Sakurai:1960, Machleidt:1987hj, Machleidt:1989tm}, 

it is assumed that the initial protons exchange a virtual meson. The interaction between the meson and the initial protons results in the production of the final state particles, which can proceed directly or via an intermediate resonance.

The exchange of a virtual meson can be put into one of two categories. The first category is strange meson exchange, where strangeness exchange occurs, and no resonances are involved. In this case, the reaction amplitude $\mathrm{KN \rightarrow KN}$ is governed by t-channel diagrams. The second category is non-strange meson exchange, a pion exchange in its simplest form. At the same time the elementary reaction amplitude $\mathrm{\pi N \rightarrow KY}$ is dominated by resonance excitations, which implies a strong and characteristic energy dependence, where Y stands for hyperons (\Lzero, \Szero, ...).

Several experiments have studied the exclusive reaction \\ \LSignal and proven that a pure phase space model description of the data is not sufficient without taking the dynamics of the process into account \cite{AbdelBary:2010pc, Munzer:2017hbl, TOF:2010ygk, Sibirtsev:2005mv}. It was found that the \Lzero hyperon production is dominated by the excitation and subsequent decay of $\mathrm{N^{*}}$ resonances to the $\mathrm{K^{+}\Lambda}$ decay channel. In particular $\mathrm{N^{*}(1650)}$ (\spin = $\frac{1}{2}^{-}$), $\mathrm{N^{*}(1710)}$ (\spin = $\frac{1}{2}^{+}$) and $\mathrm{N^{*}(1720)}$ (\spin = $\frac{3}{2}^{+}$) were found to contribute. This supports a picture wherein the exchange of non-strange mesons is the leading process in the production mechanism. In addition, a considerable Final State Interaction (FSI) was found to contribute \cite{Budzanowski:2010ib, COSY-TOF:2013uqx} leading to $\mathrm{\Sigma N \rightarrow \Lambda N}$ conversion that is observed as a $\mathrm{\Sigma N}$ cusp effect in the \Lzero cross section \cite{AbdElSamad:139102}. In the $\mathrm{pp \rightarrow pK^{+}\Sigma^{0}}$ reaction the proton–hyperon FSI seems to be negligible, especially at low energies near threshold and a pure phase space distribution describes the data reasonably well. The cross section ratio $\mathrm{\sigma(pK^{+}\Lambda)}$ / $\mathrm{\sigma(pK^{+}\Sigma^{0})}$ below excess energies of $\sim$ 20 MeV is about 28 in agreement of the SU(6) prediction and reduces drastically to about 2.5 at excess energies above 300 MeV \cite{Kowina:2004kr, Rozek:2006ct}. This energy-dependence of the cross section ratio is strongly affected by FSI effects in the \LSignal reaction \cite{Sibirtsev:54722}.

Besides the energy dependence of the cross section, the differential cross sections at selected energies add much more stringent tests of the model descriptions. This study fills this gap and delivers such data which allow some clues about the involved exchange mesons and resonances, in particular by employing a partial wave analysis.

Furthermore, a theoretical study of the reaction \SSignal based on a chiral dynamical study has been proposed in \cite{Xie:2011me}. This approach uses the pion and kaon exchange mechanisms and chiral amplitudes in addition to all pairs FSI, where the contribution of nucleon resonances appear naturally using chiral unitary amplitudes. 

This paper is organized as follows. In \cref{sec:experiment}, the experimental setup is briefly explained. \cref{sec:reconstruction} is devoted to the \Szero selection method, where the particle identification, the \Lzero hyperon reconstruction and the kinematic refit methods were presented. In \cref{sec:discussion} the method for efficiency correction and differential analysis is described. Sections \ref{xsection} and \ref{section:pwa} presents the calculated total production cross section and the partial wave analysis of the exclusive reaction \SSignal. In \cref{sec:summary} a summary and a short outlook are given.

\section{The HADES experiment}\label{sec:experiment}
The data presented here were collected in April 2007 with the High Acceptance Di-Electron Spectrometer (HADES) located at the heavy ion synchrotron SIS18 at GSI Helmholtzzentrum für Schwerionenforschung in Darmstadt, Germany. HADES is characterized by six identical sectors covering almost the full azimuthal range and polar angles from $\mathrm{\theta}$ = 18$^{\circ}$ to $\mathrm{\theta}$ = 85$^{\circ}$. 
Each sector of the spectrometer contains a Ring-Imaging Cherenkov Detector (RICH) operating in a magnetic field-free region that allows lepton identification over a wide range of momenta. Two Multi-Wire Drift Chambers (MDCs) are placed in front of a toroidal magnetic field, and two outer MDCs are placed behind the magnetic field. The MDCs enable the momentum information and the specific energy loss dE/dx to be reconstructed for each particle track. Two scintillator hodoscopes, the Time Of Flight (TOF) and TOFino are also placed behind the magnet and provide a stop time ($\mathrm{t_{s}}$) signal. The TOF and TOFino system are used as input to the trigger systems to start the data readout. A detailed description of the HADES setup can be found in \cite{HADES:2009aat}.

In the present analysis, a proton beam with an intensity of $10^{7}$ particles/s and kinetic energy T = 3.5 GeV was incident on a liquid hydrogen target with an areal density of 0.35 $\mathrm{g/cm^{2}}$. The dimensions of the target were 15 mm in diameter and 50 mm length located between -65 to -15 mm in the longitudinal direction. The data readout was started by a first level trigger requiring a charged particle multiplicity $\geq$ 3 (M3). In total, $1.14 \times 10^{9}$ events were recorded under these conditions \cite{Adamczewski-Musch:2016vrc}.

During this experiment HADES included an additional Forward Wall (FW) scintillator hodoscope that was placed 7 meters downstream the target in a magnetic field-free region and covered polar angles from $\theta$ = 0.33$^{\circ}$ to $\theta$ = 7.17$^{\circ}$ with full azimuthal acceptance. The FW measured the hit position and arrival time of the particle track with a time resolution of about 700 ps \cite{Agakishiev:2014dha}.

\section{Event selection method}\label{sec:reconstruction}
In this section, the exclusive reconstruction of the reaction \\ \SSignal is presented. The \Szero hyperon is reconstructed via its electromagnetic decay $\mathrm{\Sigma^{0} \rightarrow \Lambda \gamma}$ (BR $\approx$ 100\%) and the daughter \Lzero hyperon is reconstructed with the decay mode \\ $\mathrm{\Lambda \rightarrow p \pi^{-}}$ (BR = 63.9\%).

The \Szero reconstruction strategy includes the following steps:
\begin{enumerate*}[label=\emph{\alph*)}]
    \item time of flight ($tof$) reconstruction,
    \item charged particle identification (PID),
    \item the \Lzero hyperon reconstruction, and
    \item the \Szero hyperon reconstruction.
\end{enumerate*}

\subsection{Time of flight reconstruction}\label{subsec:tof}
The interaction of the high intensity proton beam with the START detector induced a background and prevented a stable operation of the RICH detector. Therefore, it was not possible to use the START detector information during this experiment. Consequently, the $tof$ of particle tracks were not directly measured since there was no common start time ($\mathrm{t_{0}}$) reference for tracks in the same event. The start time has to be reconstructed in order to obtain a proper time of flight measurement. 

The reconstruction algorithm is based on the assumption that at least one particle has been correctly identified. Since pions are abundantly produced, it is assumed that any negatively charged particle track that is geometrically uncorrelated to a ring in the RICH detector is a \pim. The common start time for each event is calculated by

\begin{equation*}
   t_{0} = t_{s} - \frac{d}{c}  \times \frac{\sqrt{p^{2} + m^{2}_{\pi}}}{p} \:,
\end{equation*}

\noindent
where $\mathrm{t_{s}}$ is the stop time of the \pim, $d$ is the distance to the TOF or TOFino hit, $m_{\pi}$ is the pion mass, $p$ is the momentum of the $\pi^{-}$ and $c$ is the velocity of light. The $tof$ of the other particles in the same event is the difference between the measured stop time $\mathrm{t_{s}}$ and the common start time $\mathrm{t_{0}}$. 

\subsection{Particle identification (PID)}\label{subsec:pid}
The reconstruction of the exclusive reaction $\mathrm{pp \rightarrow pK^{+}p\pi^{-}\gamma}$ only requires the identification of three particle species, pions (\pim), kaons (\Kp) and protons (\prot), since the event is kinematically complete even without measuring the photon ($\gamma$). 

As mentioned in the previous section, the \pim is identified as any negatively charged track that is geometrically uncorrelated to a ring in the RICH detector. Therefore, the problem reduces to identifying the positively charged tracks.

In order to minimize systematic bias in the model output, an auto-encoder \cite{liou2014autoencoder} implemented in PyTorch framework \cite{NEURIPS2019_9015} is trained simultaneously with both simulated and real events \cite{Esmail:2022fdy}. 
The input features used to train the auto-encoder are the momentum components, the energy loss dE/dx in the MDC and TOF sub-systems, the reconstructed $tof$ and the distance to the TOF/TOFino hit.  

A classification layer has been stacked on top of the bottleneck layer of the auto-encoder, which has three output nodes corresponding to the three classes (\pip, \Kp and \prot). Each node outputs a number between 0 and 1, where all output numbers sum to 1, so that each number can be interpreted as a probability of being a specific particle species. The network is trained by minimizing a cost function that is defined as the binary cross-entropy loss \cite{bce_good}. Because the network outputs three probabilities for each particle track, the node with the largest probability is chosen.

The classification accuracy evaluated on a holdout data-set is 92\% for pions, 76\% for kaons and 98\% for protons. It is much lower in the case of kaons since their production rate is suppressed with respect to the protons and pions.

\subsection{\Lzero hyperon reconstruction}\label{subsec:lambdareco}
The next step after the PID is to reconstruct the intermediate \Lzero hyperon. In this analysis the \Lzero reconstruction method is two-fold. In the first case, which is referred to as the \textit{Spectrometer data-set}, events with exactly 2 protons, 1 pion and 1 kaon are required to be within the main HADES detector acceptance. The other case, referred as the \textit{WALL data-set}, events were accepted if exactly 1 proton, 1 pion and 1 kaon are registered in HADES and in addition one hit in the FW. In the latter case, it is assumed that the hit registered in the FW is due to the daughter proton from the \Lzero decay (marked as $\mathrm{p_{decay}}$).

A common primary vertex in each event is then defined as the intersection point or the Point of Closest Approach (PCA) of the proton and kaon tracks. Since there is more than one proton in each event in the \textit{Spectrometer data-set}, the proton-kaon pair with the smaller Distance of Closest Approach (DCA) is used to define the primary vertex. To reduce the contribution from off-target events, a two dimensional selection is applied on the primary vertex position $(x,y,z)$:

\begin{enumerate}[label=\emph{\alph*)}]
    \item -65 $<$ $z$ [mm] $<$ -15 and \\
    \item $\sqrt{x^{2}+y^{2}}$ $<$ 5 [mm].
\end{enumerate}

\subsubsection*{The Spectrometer data-set}

\begin{figure*}[t]
    \centering
    \includegraphics[width=1.0\linewidth]{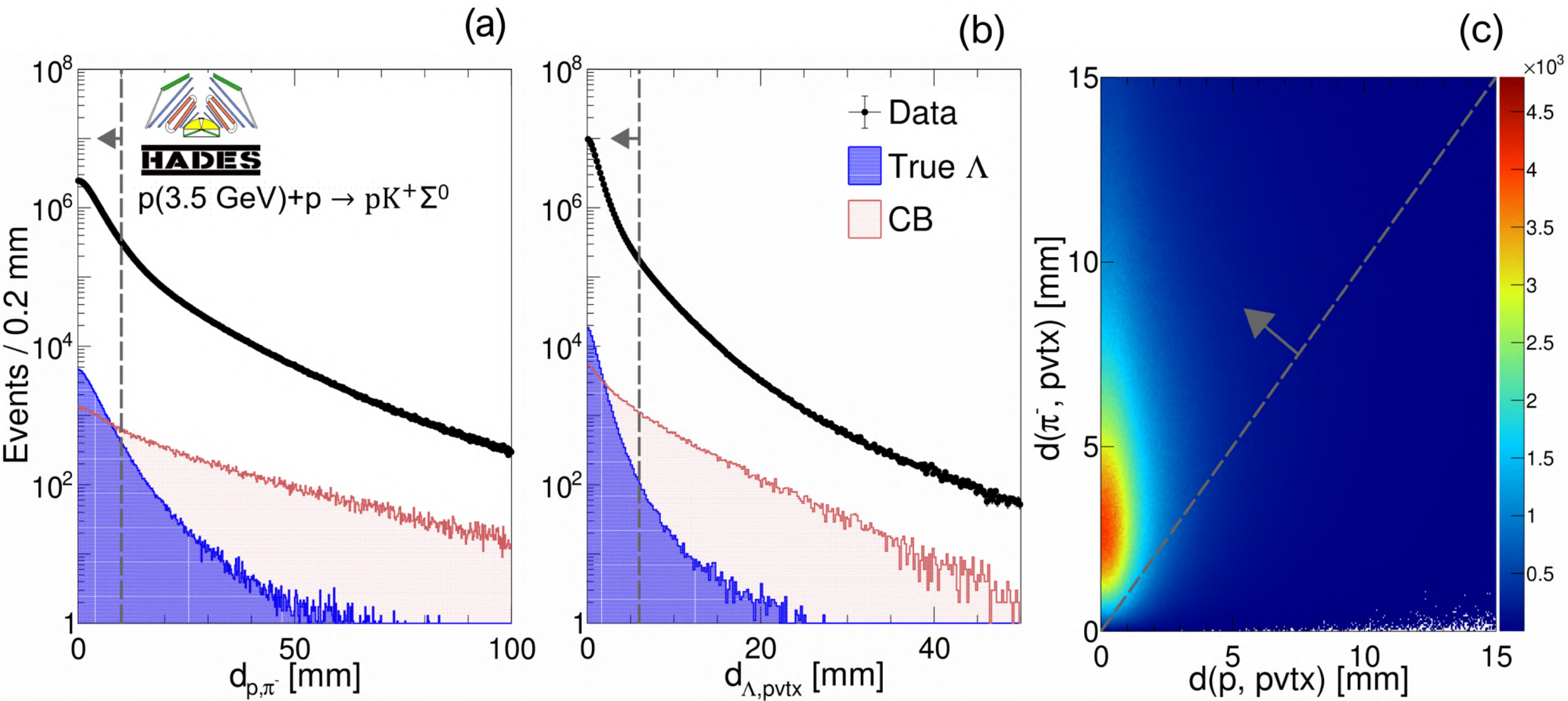}
    \caption{(\textcolor{red}{a}) The DCA distribution between the \prot and \pim tracks. (\textcolor{red}{b}) The DCA distribution between the \Lzero track and the primary vertex. In both panels, data are shown by the black points, the blue histogram represents the true \Lzero and the red histogram represents the CB, where both the true \Lzero and the CB were estimated from the simulation. (\textcolor{red}{c}) Distribution of the DCA between the \pim track and the primary vertex as a function of the DCA between the \prot track and the primary vertex. The arrows indicate the accepted regions.}
    \label{figure:hades_lambda_cuts}
\end{figure*}

Since the daughter \Lzero decays weakly ($c\tau$ = 7.89 cm), it can be identified by its displaced vertex. First, all possible combinations of the two \prot and \pim candidates were made, leaving the decision about which is the decay proton (\Lzero $\rightarrow$ \prot \pim) for later. For each combination the decay vertex (the displaced vertex) is defined as the PCA between the two tracks. The DCA between the \prot and \pim tracks (marked as $\mathrm{d_{p \pi^{-}}}$) is expected to be small if the tracks stem from the same vertex. Therefore, an upper limit of $\mathrm{d_{p \pi^{-}}}$ \textless\xspace 10 mm is imposed in order to reduce Combinatorial Background (CB), which originates from combining the wrong \prot and \pim pairs. Considering momentum and energy conservation, the \prot should be emitted in nearly the same direction as the \Lzero in the laboratory reference frame, while the \pim will have a different direction. Thus, the DCA between the \prot track and the primary vertex ($\mathrm{d_{p,pvtx}}$) is required to be smaller than the DCA between the \pim track and the primary vertex ($\mathrm{d_{\pi^{-},pvtx}}$). Finally, the DCA between the calculated \Lzero track and the primary vertex ($\mathrm{d_{\Lambda,pvtx}}$) is required to be \textless\xspace 6 mm. The distributions of the topological variables are shown in \Cref{figure:hades_lambda_cuts}, where the selection criteria are indicated by the vertical dashed lines. The proton used in the \Lzero reconstruction is tagged as the decay proton (marked in the following as \pdecay), while the other proton in the event is tagged as the scattered (primary) proton.

\begin{figure}[!hbt]
    \centering
    \includegraphics[width=1.0\linewidth]{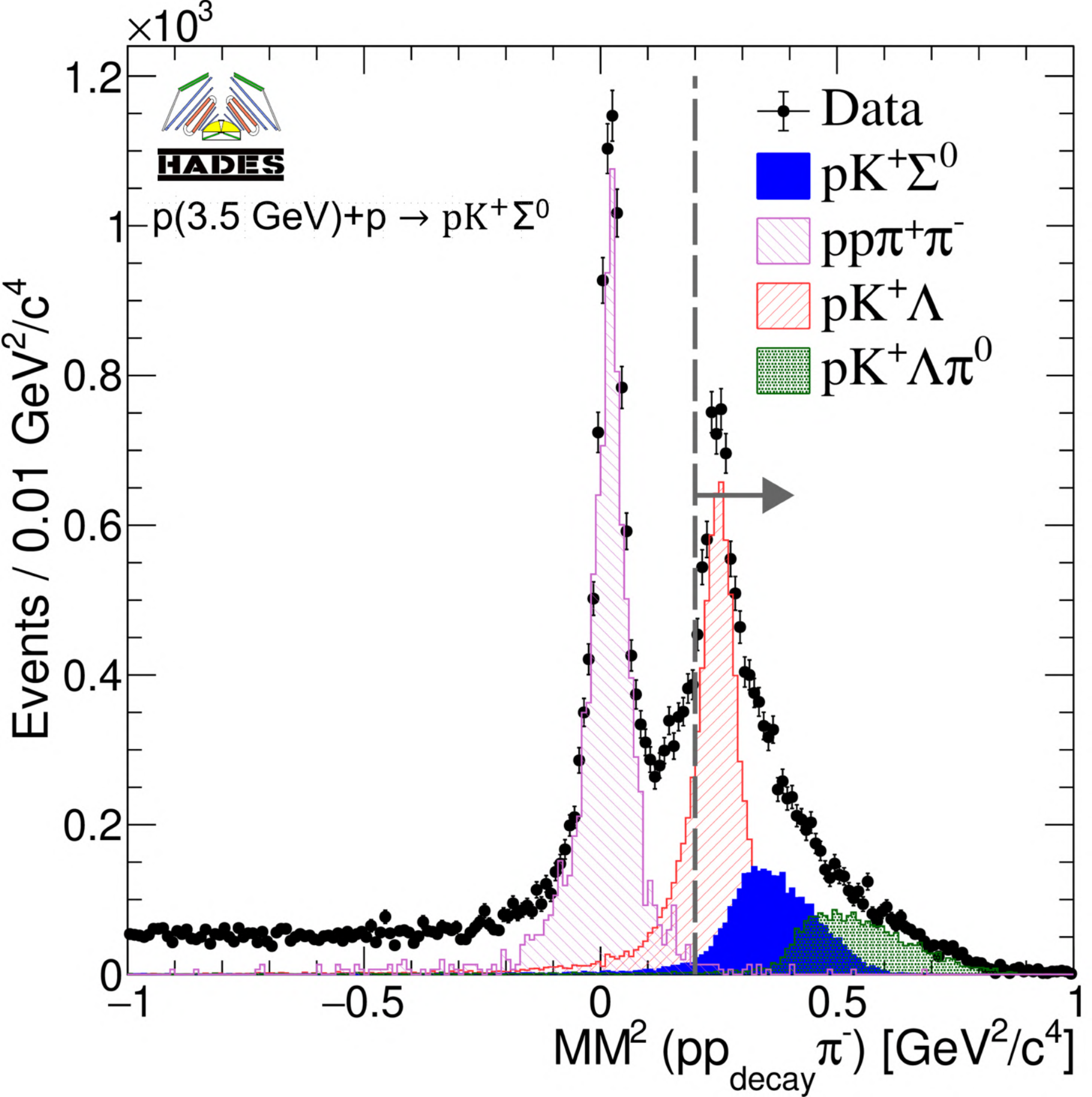}
    \caption{The squared p\pdecay\pim missing mass distribution after applying the topological selections. Black points are the \textit{Spectrometer data-set} data. The violet histogram is the $\mathrm{pp \rightarrow p p \pi^{+} \pi^{-}}$ simulation. The \LSignal, \SSignal and \LBack simulations are shown by the red, blue and green histograms, respectively. The vertical line and the arrow indicate the accepted region for the further analysis.}
    \label{figure:hades_mm2}
\end{figure}

To further purify the selected \Lzero sample, the event kinematics were constrained to the \Szero production range. The squared $\mathrm{p\Lambda}$ missing mass ($\mathrm{MM^{2}(pp_{decay}\pi^{-})}$) is required to be \textgreater\xspace 0.2 \GeV in order to reject the multi-pion production channel as shown in \Cref{figure:hades_mm2}. In this figure, the experimental data are shown by the black points and the simulations (discussed in \cref{sec:hyperon}) by different colored histograms. Two peaks are visible, the first peak at 0.02 \GeV corresponds to the multi-pion channel via the reaction $\mathrm{pp \rightarrow p p \pi^{+} \pi^{-}}$ (violet histogram), where a $\mathrm{p\pi^{-}}$ pair is incorrectly identified as a \Lzero candidate and the \pip is incorrectly identified as a \Kp. The other broader peak is the sum of \LSignal, \SSignal and \LBack reactions shown by the red, blue and green histograms, respectively. The relative normalizations of the simulated channels have been chosen to best fit the experimental data as explained in \Cref{sec:scaling}.

\begin{figure}[!hbt]
    \centering
    \includegraphics[width=1.0\linewidth]{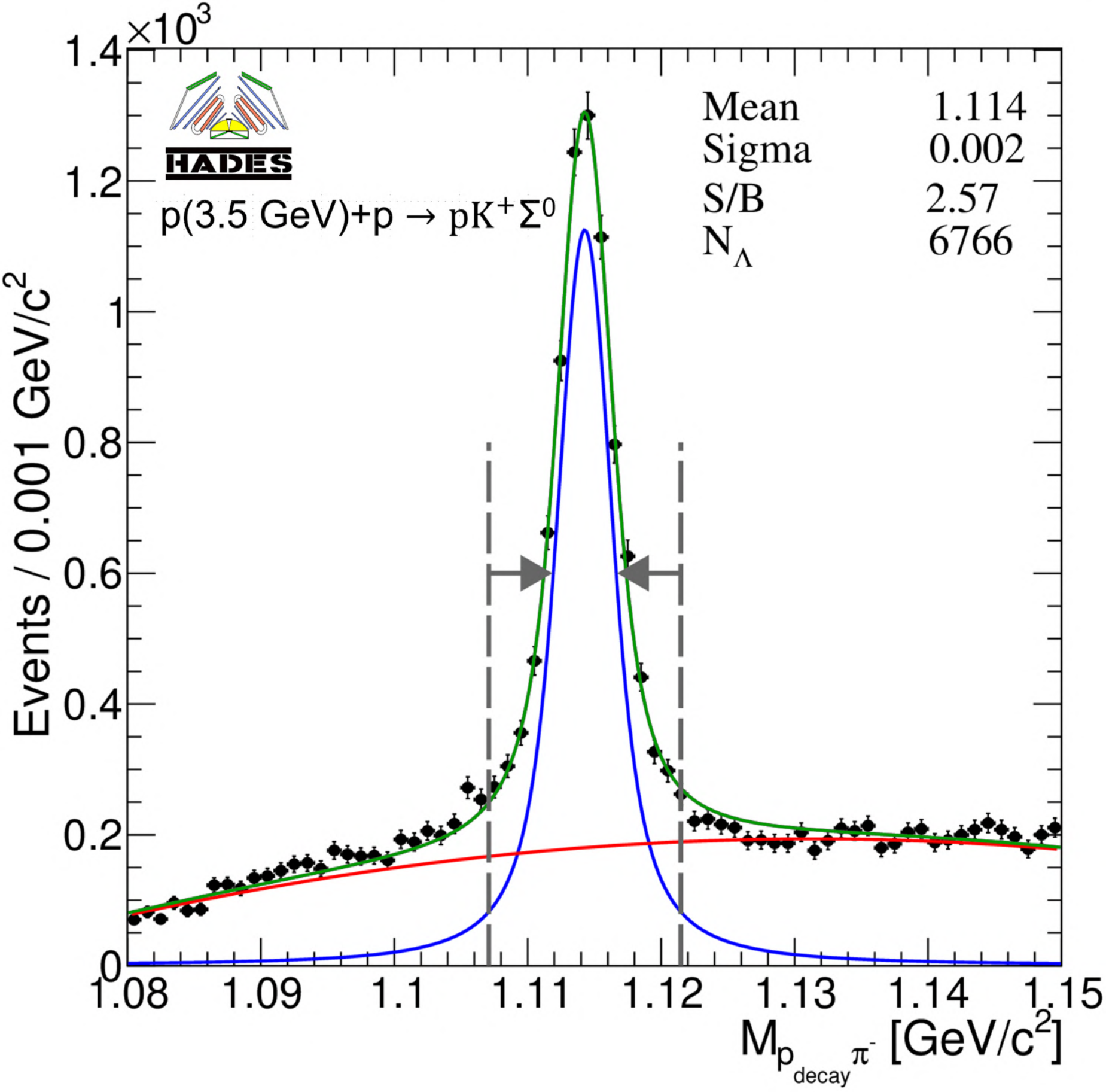}
    \caption{The \pdecay\pim invariant mass distribution. The vertical dashed lines indicate the selected mass range. The blue, red and green curves are for the signal, background and the total fit.}
    \label{figure:hades_im}
\end{figure}

The \pdecay\pim invariant mass distribution is shown in \Cref{figure:hades_im}. A peak around the nominal \Lzero mass is visible on top of background. The signal has been parameterized by a Voigt distribution and the background is modeled by a fourth-order polynomial. Events are further processed if they are in the range of $\mathrm{\mu \pm 3\sigma}$, where the calculated signal to background ratio in this range is $\mathrm{S/B = 2.57}$ and the number of \Lzero candidates is N$_{\mathrm{\Lambda}}$ = 6766.

\subsubsection*{The WALL data-set}
\begin{figure*}[t]
    \centering
    \includegraphics[width=1.0\linewidth]{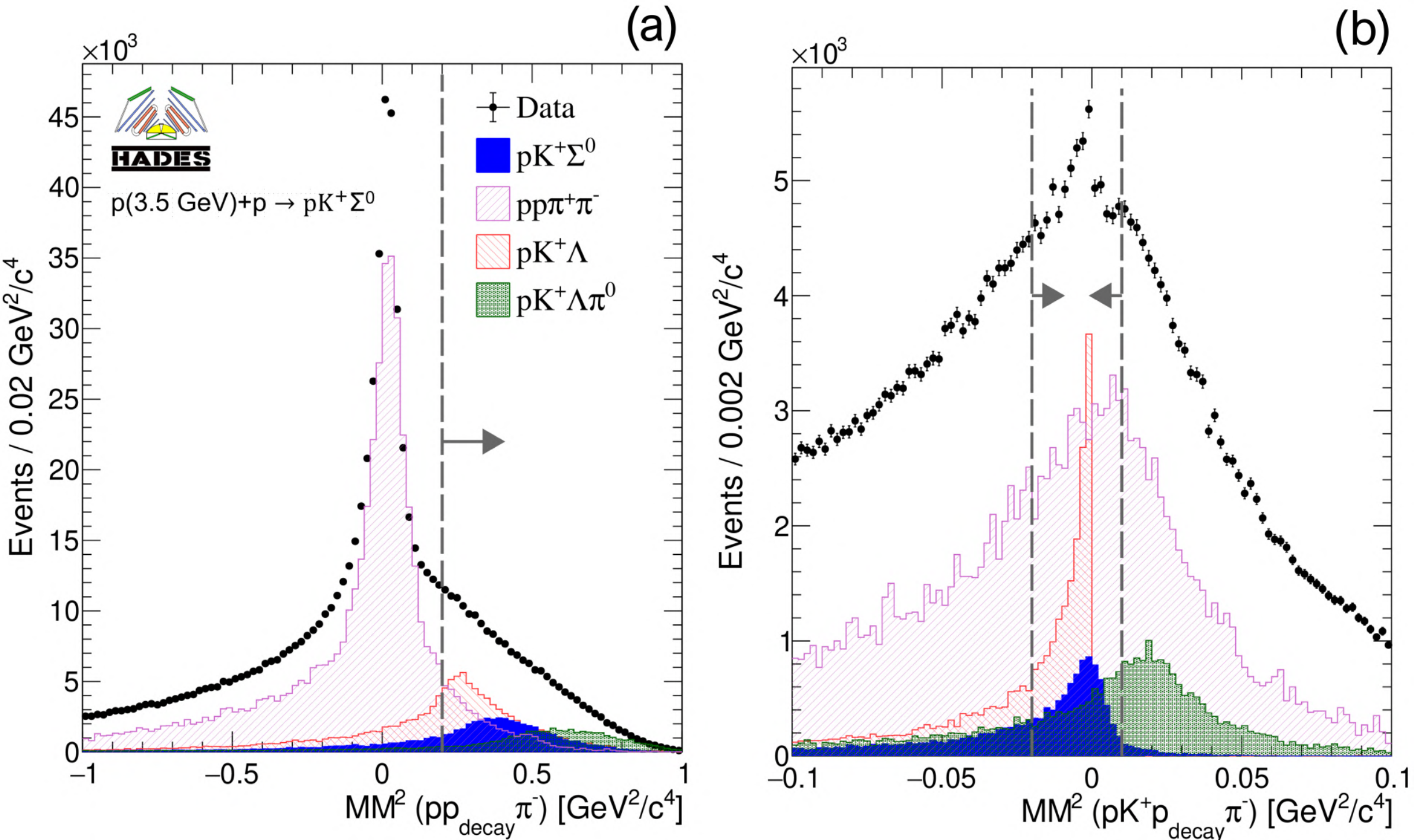}
    \caption{(\textcolor{red}{a}) The squared p\pdecay\pim missing mass distribution of \textit{WALL data-set}. (\textcolor{red}{b}) The squared p\pdecay\pim\Kp missing mass distributions. The $\mathrm{pp \rightarrow p p \pi^{+} \pi^{-}}$, \LSignal, \SSignal and \LBack simulations are shown by the violet, red, blue and green histograms, respectively. The arrows indicate the accepted regions.}
    \label{figure:wall_mm}
\end{figure*}

In the \textit{WALL data-set} the hit in the FW is assumed to be due to the decay proton. Since the FW is installed in a magnetic field-free region, the \pdecay is reconstructed as a straight line trajectory from the primary vertex position to the hit in the FW. The track momentum is calculated from the $tof$ and the distance from the primary vertex and the FW detector hit, assuming the proton mass. In this case, the topological cuts are not as effective to suppress the background as in the \textit{Spectrometer data-set}. Therefore, events fulfilling the following kinematical constraints were selected:

\begin{enumerate}[label=(\roman*)]
    \item $\mathrm{MM^{2}(pp_{decay}\pi^{-})}$ \textgreater\xspace 0.2 \GeV (\Cref{figure:wall_mm}\xspace\textcolor{red}{a}) and \\
    \item The squared missing mass of all charged particles is required to be in the following range: \\ $\mathrm{-0.02 < MM^{2}(pK^{+}p_{decay}\pi^{-}) [GeV^{2}/c^{4}] < 0.01}$ because \\ only a photon is missing to completely measure the exclusive final state (\Cref{figure:wall_mm}\xspace\textcolor{red}{b}).
\end{enumerate}

\begin{figure}[t]
    \centering
    \includegraphics[width=1.0\linewidth]{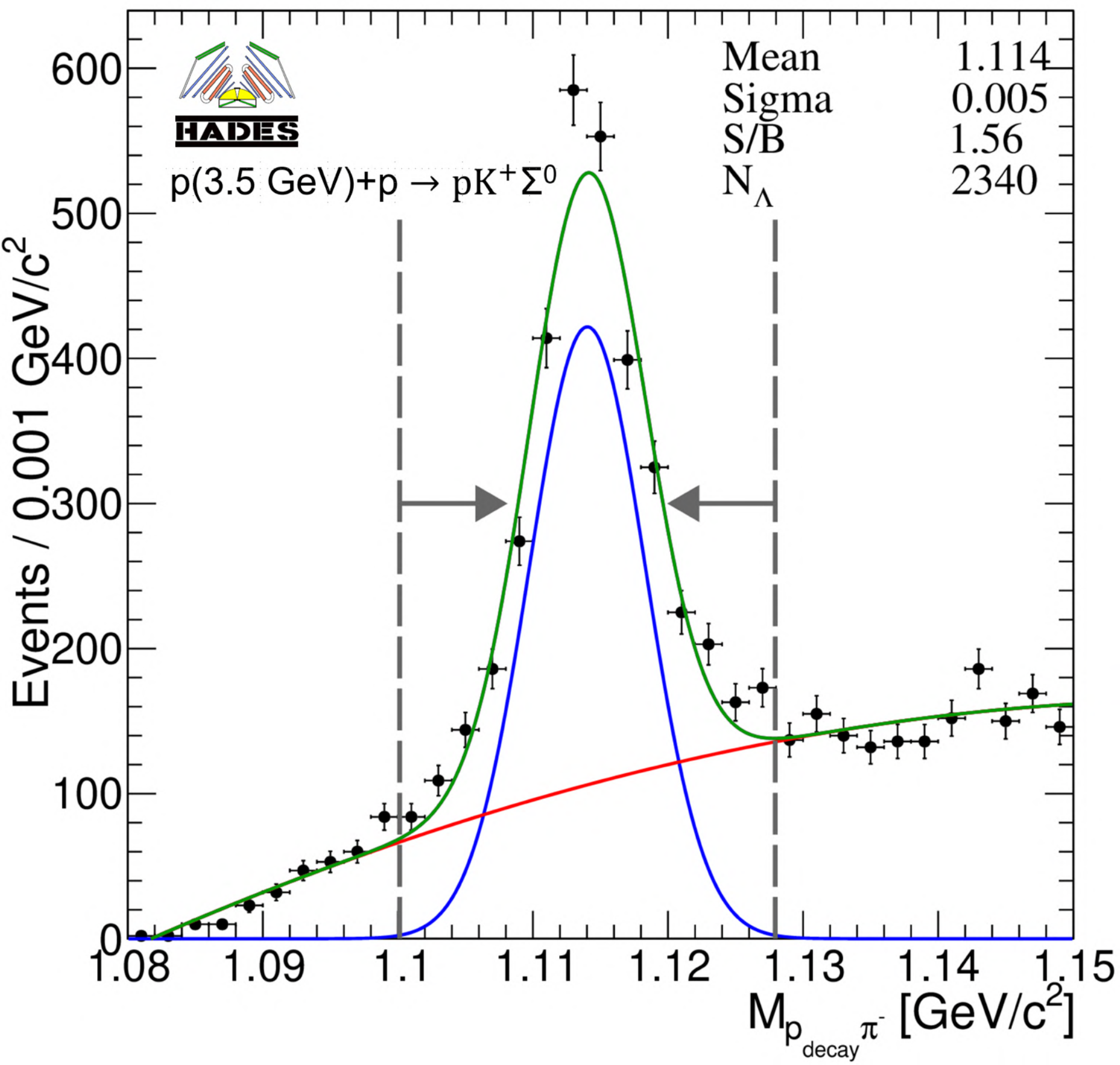}
    \caption{The $\mathrm{p_{wall}\pi^{-}}$ invariant mass distribution. The vertical dashed lines indicate the selected mass range. The blue, red and green curves are for the signal, background and the total fit.}
    \label{figure:wall_im}
\end{figure}

The $\mathrm{p_{decay}\pi^{-}}$ invariant mass distribution for the \textit{WALL data-set} is shown \Cref{figure:wall_im} after applying the selections mentioned above. Once again, the peak has been fitted by a Voigt distribution and the background by a fourth-order polynomial. However, the mass resolution of the \Lzero peak of the \textit{Spectrometer data-set}(\Cref{figure:hades_im}) is better than the signal of the \textit{WALL data-set}, since in the latter case the proton was detected in the FW, which has a worse momentum resolution. Events are further processed if they are in the range of $\mathrm{\mu \pm 3\sigma}$, where the calculated signal to background ratio in this range is $\mathrm{S/B = 1.56}$ and the number of \Lzero candidates is N$_{\mathrm{\Lambda}}$ = 2340.

\subsection{\Szero hyperon reconstruction}\label{sec:hyperon}
To further suppress the remaining background and to obtain a better mass resolution, a kinematic fit based on the Lagrange multiplier method is employed \cite{osti_837857}. The fit $\mathrm{\chi^{2}}$, expressed as

\begin{equation*}
    \chi^{2}(\eta,\lambda) = (y-\eta)^{T}V(y)(y-\eta) + 2\lambda^{T}f(\eta)\:,
\end{equation*}

\noindent
is minimized by differentiating $\chi^{2}$ with respect to all measured variables. Here $y$ is a vector containing the initial guesses for the measured quantities, which are the track parameters provided by the tracking algorithm, $\eta$ is an improved set of the track parameters and $V$ is the covariance matrix comprising the estimated errors on the measured quantities. The constraint equations are expressed as a function of $\eta$ in $f(\eta)$, where $\lambda_{i}$ are a set of Lagrange multipliers.

The spherical coordinates used in this analysis for the track parameterization are defined as follows

\begin{equation*}
  y = 
  \begin{bmatrix}
    1/p \\
    \theta \\
    \phi
  \end{bmatrix} \:,
\end{equation*}

\noindent
where $1/p$ is the inverse of the absolute momentum, $\theta$ and $\phi$ are the polar and azimthual angles of the track.

Two constraints were applied to both data-sets. The first is the proton and pion from the \Lzero decay are constrained to the \Lzero mass ($\mathrm{M_{\Lambda}} = $ 1.1157 \GeV). The second constraint is that the missing mass of all the charged final state particles is constrained to the photon mass ($\mathrm{M_{\gamma}} = $ 0 \GeV).

\begin{figure*}[t]
    \centering
    \includegraphics[width=1.0\linewidth]{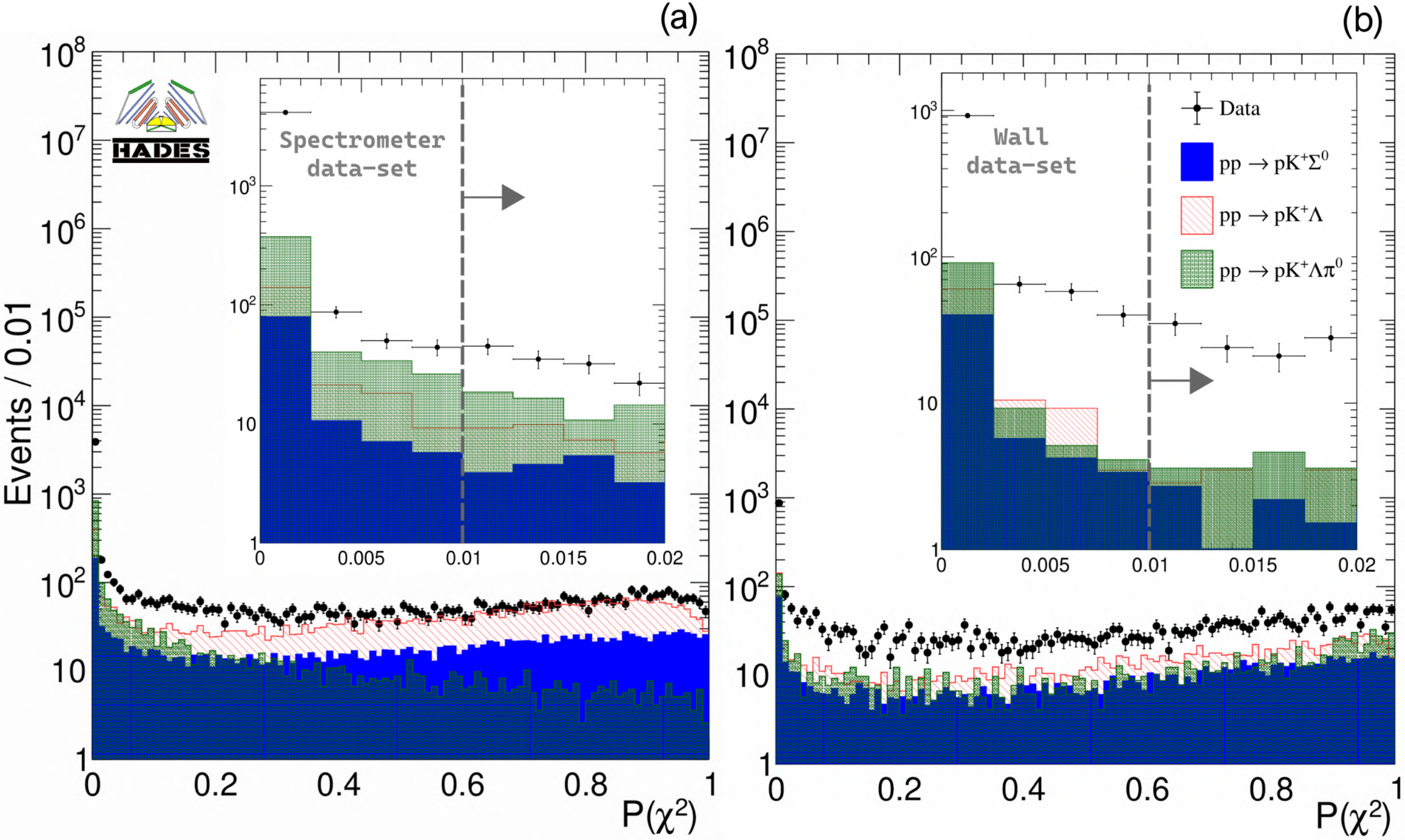}
    \caption{(\textcolor{red}{a}) The p-value distributions for the HADES data-set and for (\textcolor{red}{b}) the WALL data-set. The insets display the region of small p-values, where the dashed line and the arrow indicates the accepted region. The \LSignal, \SSignal and \LBack simulations are shown by the red, blue and green histograms, respectively.}
    \label{figure:kfit}
\end{figure*}

The probability that a $\mathrm{\chi^{2}}$ of the theoretical distribution is greater than or equal to the $\mathrm{\chi^{2}}$ value found from the fit is known as the p-value ($\mathrm{P(\chi^{2})}$). The p-value distributions of the \textit{Spectrometer} and the \textit{WALL} data-sets are shown in \Cref{figure:kfit}. Because both \Lzero and \Szero have $\mathrm{MM(pK^{+}\Lambda)}$ = 0, they have similar distributions, which makes these two reactions difficult to distinguish. On the other hand, the reaction \LBack should ideally have zero p-value. However, due to the limited detector resolution it has p-values greater than zero, which is more pronounced in the \textit{WALL} data-set. The signal events show an almost flat distribution between 0 and 1, while events that do not satisfy the constraint equations have a prominent yield of p-values close to 0. Therefore, events with p-values \textgreater\xspace 0.01 are selected, where the cut was optimized based on a significance analysis.

\subsubsection*{Simulation scaling to the experimental data}\label{sec:scaling}
\begin{figure}[!hbt]
    \centering
    \includegraphics[width=1.0\linewidth]{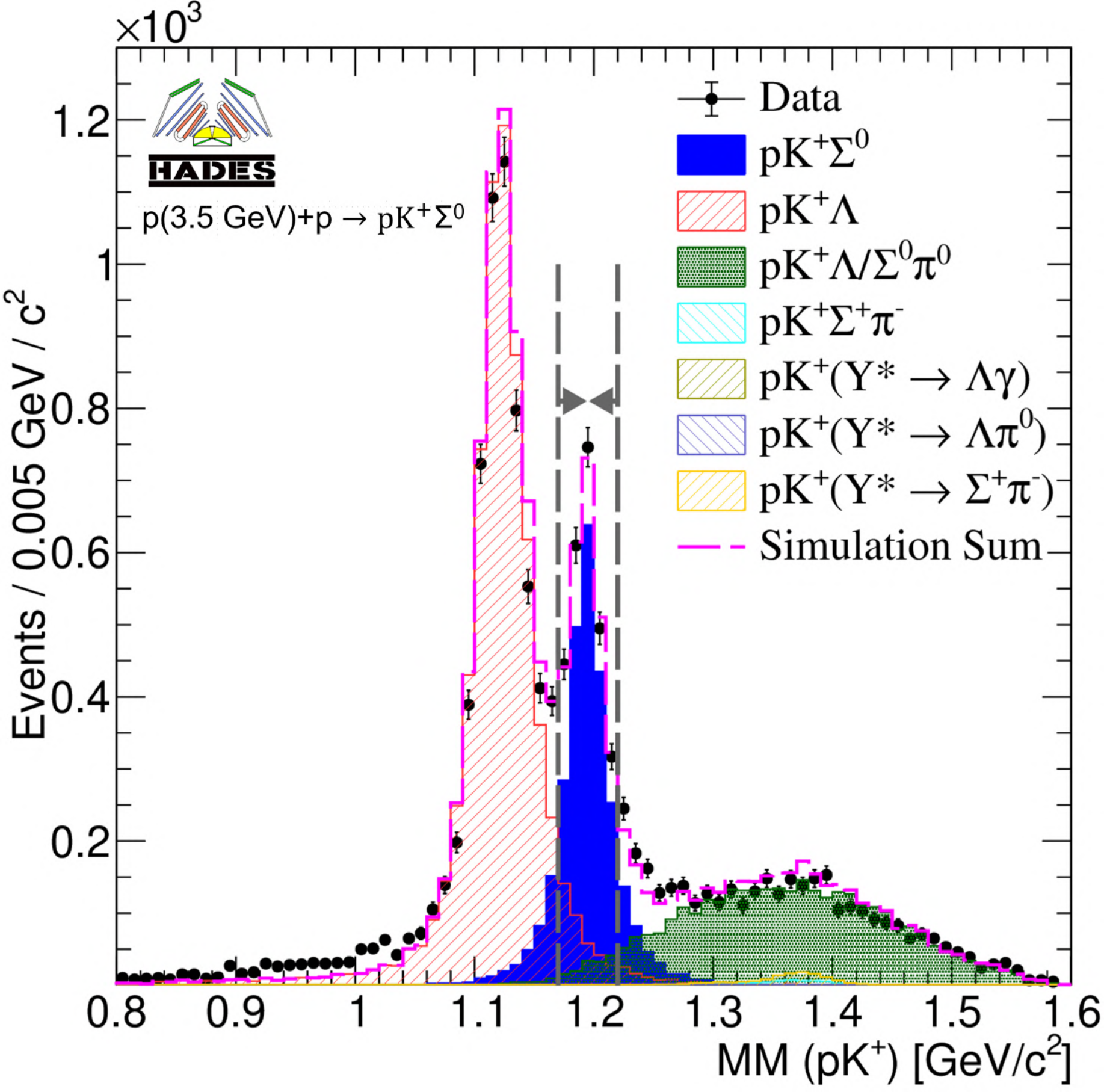}
    \caption{The $\mathrm{pK^{+}}$ missing mass distribution. The colored histograms represent the simulated channels, where Y$^{*}$ refers to an excited hyperon (\Sstar, \Lstar or \Lstard). The two peaks are due to the exclusive reactions \LSignal and \SSignal as shown by the red and the blue histograms, respectively. The vertical dashed lines mark the mass window used to select candidate events of the \SSignal final state.}
    \label{figure:combined_mm}
\end{figure}

By inspecting the $\mathrm{pK^{+}}$ missing mass distribution of the combined data-set shown in \Cref{figure:combined_mm}, two peaks corresponding to the \Lzero and the \Szero, as well as other minor contributions in the high mass region, are plainly evident. In order to quantify the different contributions an incoherent cocktail has been simulated using the Pluto event generator \cite{Frohlich:2007bi}. All the simulated reactions have been processed using the same full scale analysis employed for the experimental data, thus taking into account the efficiency of the trigger condition, the tracking algorithm and the analysis procedure. The particle decays, the acceptance and the particle interactions with the materials of HADES and the FW have been considered by using GEANT3 \cite{Brun:1994aa}.

To determine the contributions of the different channels, a fit of the simulations to the measured missing mass spectrum ($\mathrm{MM(pK^{+})}$) has been carried out by minimizing the quantity

\begin{align*}
    \chi^{2} = \sum_{i}^{n_{bins}} \frac{(n_{data} - \sum_{ch} (f^{ch} \times n^{ch}_{simulation}))^{2}}{\sigma_{data}^{2} + \sigma_{simulation}^{2}} \:,
\end{align*}

\noindent
where the summation runs over the number of bins of the missing mass spectrum, $\mathrm{n_{data}}$ is the number of data events in each bin, $\mathrm{n^{ch}_{simulation}}$ is the number of simulated events in each bin for each channel and $\mathrm{f^{ch}}$ is a scaling factor for each channel. The uncertainty for the data and the simulations in each bin is $\mathrm{\sigma_{data}}$ and $\mathrm{\sigma_{simulation}}$, respectively.

As can be seen from \Cref{figure:combined_mm}, the experimental data is primarily described by contributions of \LSignal,\\ \SSignal and \LBack indicated by the red, blue and the green histogram, respectively. The other simulated channels have minor contributions. In total 2613 \Szero candidates were collected within the $\mathrm{pK^{+}}$ missing mass range of 1.170-1.220 \GeV, 58\% of them are within the main HADES acceptance and 42\% within the FW acceptance. The signal purity in the mass window calculated from the simulation is found to be 81\%, where the main background contributions are the reactions \LSignal (14\%) and \LBack (5\%).

\subsection{Efficiency and acceptance correction}\label{sec:discussion}
The reconstructed experimental distributions are corrected for the limited detector acceptance and efficiency by using a simulated phase space distribution that were assigned a weight determined by the best partial wave solution (discussed in \cref{section:pwa}), then the events were filtered through the full scale simulation and analysis. The efficiency correction is done in one dimension whereas the other three degrees of freedom on which the efficiency depends are integrated. 

The 1D correction matrix (M) is calculated given the initial 4$\pi$ distribution (T) for each observable and after filtering through the full scale simulation and analysis (R). To put it another way, a distinct correction matrix M = R/T is constructed for each angular distribution shown in \Cref{figure:angularD}. The inverse of the correction matrix is then calculated using the Singular Value Decomposition (SVD) technique \cite{Hocker:1995kb} implemented in RooUnfold framework \cite{Adye:2011gm}.

\subsection{Absolute normalization and systematic uncertainties}
The production cross section of \Szero can be calculated by normalizing the corrected \Szero yield to the p+p elastic scattering yield measured in the same experimental run \cite{HADES:2011ab}. This normalization results in a systematic uncertainty of 7\%. In addition, there might be other sources of systematic uncertainty. The systematic error associated to the exclusive event selection has been estimated by varying the selection ranges and recalculating the cross section.

To test the influence of different selection cuts on the calculated cross section (see section \ref{xsection}), the whole analysis chain has been repeated many times under different cut combinations. Each cut is varied in two steps in either direction. the cross section for each combination is then calculated by integrating the yield of the $\mathrm{cos \theta^{*}_{\Sigma^{0}}}$ angular distribution. Following this procedure the obtained systematic error, defined as the 1$\mathrm{\sigma}$ interval of the cross sections distribution, is found to be $\approx$ 2\%.

Another source of the systematic errors is the PID, which is evaluated by activating the dropout layers of the neural network during the inference time as this is equivalent to doing a Bayesian approximation \cite{pmlr-v48-gal16}. The estimated size of the PID systematic is $\approx$ 5\%.

\section{Angular Distributions}
\begin{figure*}[t]
    \centering
    \includegraphics[width=1.0\linewidth]{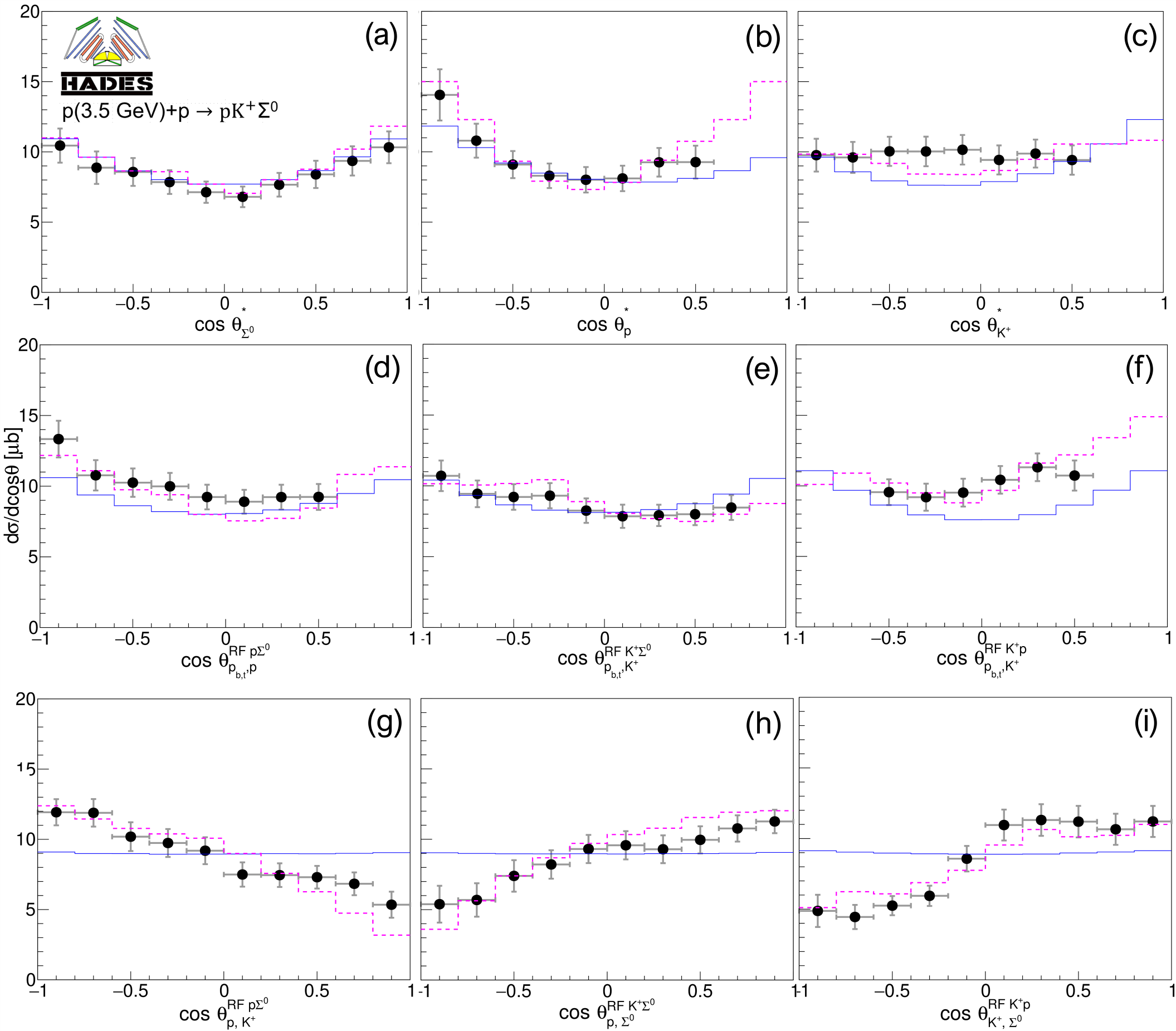}
    \caption{The corrected angular distributions in the CMS (top row), Gottfried-Jackson (middle row) and helicity frames (bottom row). The experimental data are shown by the black points, where the error bars are the square root of the quadratic sum of the statistical and systematic uncertainties. The blue histogram represent the weighted \SSignal phase space simulation described in the text and the dotted pink histogram indicates the best partial wave analysis solution (discussed in \Cref{section:pwa}).}
    \label{figure:angularD}
\end{figure*}

This section presents the differential cross section of the reaction \SSignal, namely the angular distributions of final state particles in the center-of-mass (CMS) frame, as well as in both the Gottfried-Jackson and helicity frames of all two-body subsystems. All distributions are acceptance and efficiency corrected and then fit with Legendre polynomials $d\sigma/dcos\theta$ = $\sum_{l} A_{l}\cdot P_{l}$, with $l$ = $0,1,2$. The coefficients $A_{1}$ and  $A_{2}$ are used to judge the asymmetries and anisotropies of the observed distributions. The best description of the distribution (indicated by the blue histogram in \Cref{figure:angularD}) was found when the simulations have been weighted simultaneously with the angular distribution of the \Szero hyperon in the CMS frame and the proton Gottfried-Jackson angular distribution measured in the $\mathrm{p\Sigma^{0}}$ rest frame obtained from the data.

\begin{table}[t]
 \caption{Coefficients of Legendre polynomials determined by fitting the angular distributions presented in \Cref{figure:angularD}.}
 \label{table:Legender}
 \begin{tabularx}{1.0\linewidth}{L{0.25}C{0.25}C{0.25}C{0.25}}
 \toprule
  Angle & A$_{0}$ [$\mathrm{\mu b}$] & A$_{1}$ [$\mathrm{\mu b}$] & A$_{2}$ [$\mathrm{\mu b}$] \\
 \midrule
    $\mathrm{cos\theta^{*}_{\Sigma^{0}}}$ &  8.55 $\pm$ 0.31 & 0.00 & 2.75 $\pm$ 0.73 \\
    
    $\mathrm{cos\theta^{*}_{p}}$ & 10.01 $\pm$ 0.50 & 0.00 & 4.33 $\pm$ 1.27 \\
    
    $\mathrm{cos\theta^{*}_{K^{+}}}$ &  9.83 $\pm$ 0.43 & 0.00 & -0.13 $\pm$ 1.02 \\
    
    $\mathrm{cos\theta^{FR p\Sigma^{0}}_{p_{b,t}, p}}$ & 10.40 $\pm$ 0.80  & -0.64 $\pm$ 1.73 & 2.79 $\pm$  1.85  \\
    
    $\mathrm{cos\theta^{FR K^{+}\Sigma^{0}}_{p_{b,t}, K^{+}}}$ & 8.55 $\pm$ 0.71 & -1.61 $\pm$ 1.54 & 0.66 $\pm$ 1.63 \\
    
    $\mathrm{cos\theta^{FR K^{+}p}_{p_{b,t}, K^{+}}}$ & 10.30 $\pm$ 1.00 & 1.91 $\pm$ 1.18 & 0.50 $\pm$ 2.69  \\
    
    $\mathrm{cos\theta^{FR K^{+}\Sigma^{0}}_{p, \Sigma^{0}}}$ & 8.70 $\pm$ 0.30 & 3.17 $\pm$ 0.59 & -0.73 $\pm$ 0.75  \\
     
    $\mathrm{cos\theta^{FR p\Sigma^{0}}_{p, K^{+}}}$ & 8.75 $\pm$ 0.29 & -3.52 $\pm$ 0.50 & 0.37 $\pm$ 0.67 \\
      
    $\mathrm{cos\theta^{FR K^{+}p}_{K^{+}, \Sigma^{0}}}$ & 8.81 $\pm$ 0.31 & 4.84 $\pm$ 0.56 & -0.98 $\pm$ 0.75 \\
 \bottomrule
 \end{tabularx}
\end{table}

\subsubsection*{Center of mass frame}
The angular distributions of the three final state particles in the CMS are shown in the top row of \Cref{figure:angularD}. The Legendre polynomial coefficients obtained from the fits of the angular distributions are listed in \Cref{table:Legender}. Since the initial p+p is a symmetric system, the $A_{1}$ Legendre parameters of all CMS distributions were set to zero. The angular distribution of the \Szero hyperon (\Cref{figure:angularD} (a)) and proton (\Cref{figure:angularD} (b)) shows an anisotropy, where it is more pronounced for the proton as quantified by the $\mathrm{A_{2}}$ parameter listed in \Cref{table:Legender}. From the observed anisotropies and the fit parameters one deduces that a non-zero orbital angular momentum ($L$) is observed in both the $\mathrm{p-K^{+}\Sigma^{0}}$ and $\mathrm{\Sigma^{0}-pK^{+}}$ sub-systems. This is in contrast to the kaons, where the angular distribution is compatible with isotropy. For pure pion exchange, the final state proton is the leading particle, since the exchange pion has a small mass, implying a small 4-momentum transfer so that the proton is preferably emitted in the direction of the initial protons, which could explain the anisotropy in the proton angular distribution. In this picture, the \Szero CMS angular distribution reflects the proton one, while the kaon has a broader distribution.

The angular distributions in the overall CMS are not suited to directly draw conclusions on resonant production, which proceeds as a two step process pp \textrightarrow\xspace pR, R \textrightarrow\xspace \Kp\Szero, where R stands for every kind of nucleon resonance, that can be either an isospin 1/2 $\mathrm{N^{*}}$ state or an isospin 3/2 $\mathrm{\Delta^{*}}$ state. Therefore in the following the Gottfreid-Jackson and helicity frames are presented as a more natural choice for the Lorentzian reference frames in order to study the reaction properties due to resonant production.

\subsubsection*{Gottfried-Jackson frames}
The Gottfried-Jackson (G-J) frame first introduced in \cite{Gottfried:1964nx} is the rest frame of two out of the three produced particles. In the G-J frame, the G-J angle is defined as the angle between one of the rest frame particles (e.g. the \Szero) and the initial proton $\mathrm{\theta^{RF \, K^{+}\Sigma^{0}}_{p_{b,t},\Sigma^{0}}}$, where the label RF stands for reference frame, the superscript indicates which rest frame is used and the subscript stands for the two particles, between which the angle is measured. It should be noted that the two initial protons are indistinguishable. Therefore, the angular distribution is calculated by using the angle to both protons ($\mathrm{p_{b,t}}$).

In the case of kaon (pion) exchange, the $\mathrm{K^{+}p}$ ($\mathrm{K^{+}\Sigma^{0}}$) rest frame is equivalent to the rest frame of the exchanged meson and the initial proton. In this way, the initial 2 \textrightarrow\xspace 3 reaction is reduced to a pure 2 \textrightarrow\xspace 2 reaction. If there is a resonant production, the internal angular momentum of the resonance is then reflected in this observable. It has to be noted that the distributions in the G-J frames do not have to be symmetric. The reason is the asymmetric reaction system, where either a kaon or a pion collides with a proton. The angular distributions in the G-J frames are shown in the middle row of \Cref{figure:angularD}. 

\begin{figure}[!hbt]
    \centering
    \includegraphics[width=1.0\linewidth]{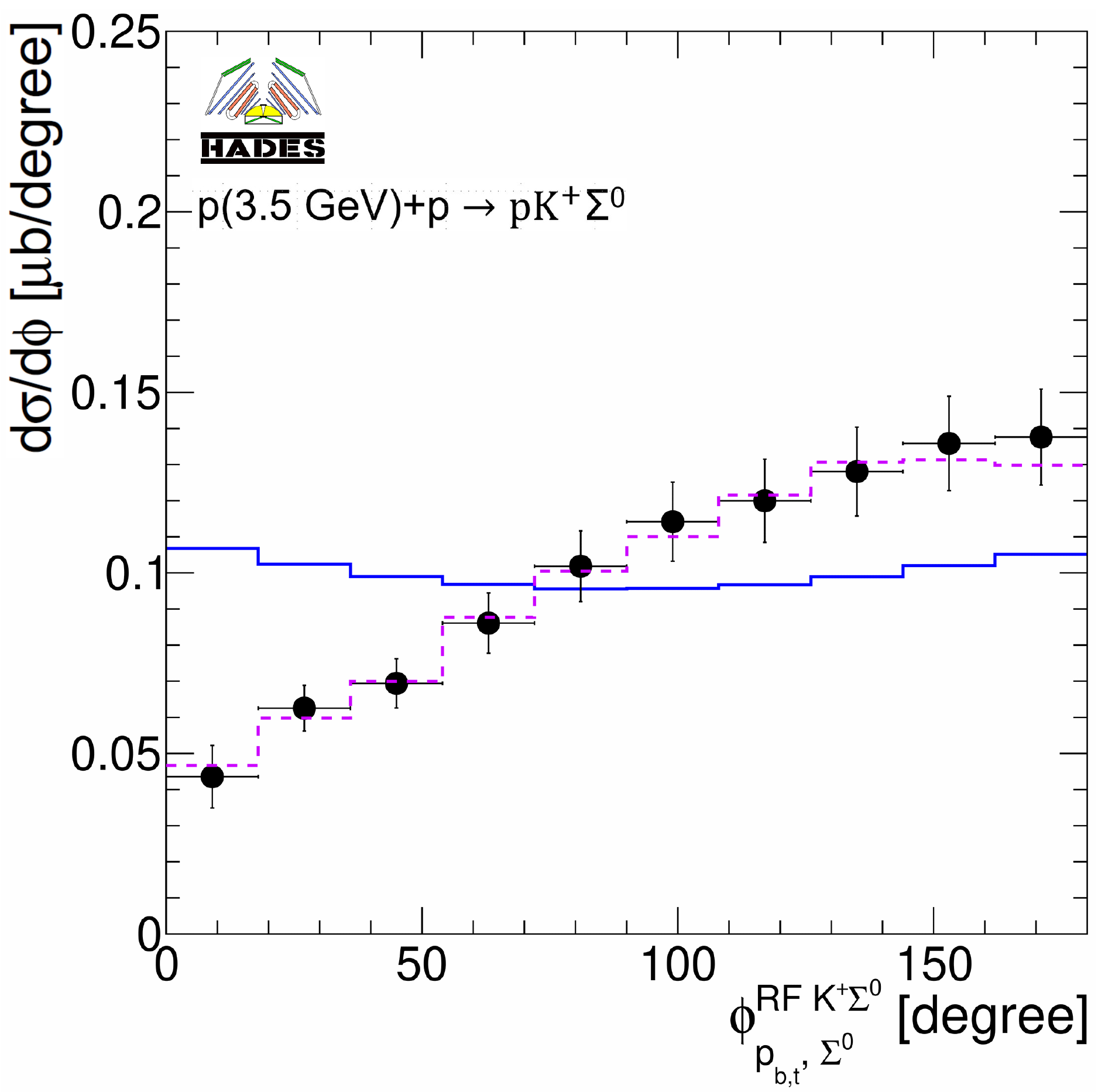}
    \caption{The \Szero Treiman-Yang angular distribution measured in the $\mathrm{K^{+}\Sigma^{0}}$ reference frame. The blue histogram represents the weighted \SSignal phase space simulation and the dotted histogram indicates the best partial wave analysis solution (discussed in \Cref{section:pwa}).}
    \label{figure:Yang}
    \end{figure}

An anisotropy is observed in the $\mathrm{p\Sigma^{0}}$ G-J frame (\Cref{figure:angularD} (d)), which could be due to a relative angular momentum in the $\mathrm{p\Sigma^{0}}$ system. This effect is related to the above mentioned anisotropies of the \prot and \Szero CMS angular distributions since they are kinematically related. The angular distribution in the in $\mathrm{K^{+}\Sigma^{0}}$ G-J frame (\Cref{figure:angularD} (e)) tends to be asymmetric, which could be caused by the excitation of nucleon resonances decaying into the $\mathrm{K^{+}\Sigma^{0}}$ channel \cite{AbdelBary:2010pc}. Many of $\mathrm{N^{*}}$ or $\mathrm{\Delta^{*}}$ resonances could contribute to the reaction. All these resonances have large widths and may also contribute through their broad tails to the reaction. The angular distribution of a true two-body resonance reaction is asymmetric only if resonances with both parities are simultaneously excited through interfering amplitudes. Hence, this distribution in the $\mathrm{K^{+}\Sigma^{0}}$ G-J frame indicates that more than one nucleon resonance with opposite parity participates in the production process \cite{AbdelBary:2010pc}. As explained earlier, the $\mathrm{K^{+}p}$ rest frame is equivalent to the rest frame of the exchanged kaon. Therefore, the deviation from isotropy in the $\mathrm{cos\theta^{FR K^{+}p}_{p_{b,t}, K^{+}}}$ angular distribution could be explained by kaon exchange component \cite{Esmail:2022fdy}. For a pure pion exchange, the Treiman-Yang (T-Y) angle measured in the $\mathrm{K^{+}\Sigma^{0}}$ rest frame is expected to be an isotropic distribution \cite{Ferrari:1968bbl}. Therefore, if a kaon exchange contributes to the production mechanism it should reflect itself in this distribution. The \Szero hyperon T-Y angle measured in the $\mathrm{K^{+}\Sigma^{0}}$ rest frame, shown in \Cref{figure:Yang}, shows a clear deviation from isotropy, which could be an indication of a significant kaon exchange contribution to the reaction mechanism.

\subsubsection*{Helicity frames}
The helicity angle is defined in a similar way as the G-J angle, but instead of calculating the angle of the respective particle to the initial proton, the helicity angle is calculated between one of the rest frame particles and the third produced particle. The helicity angular distribution thus interrelates the kinematics of the three final state particles and it is thus a linear transformation projection of the Dalitz plot. A uniformly populated Dalitz plot results in isotropic helicity angle distributions. On the other hand, if dynamical effects distort the Dalitz plot, then the helicity angular distribution will be anisotropic. The helicity angular distributions are shown in the bottom row of \Cref{figure:angularD}. All the distributions are significantly non-isotropic, which indicates that the reaction is dominated by intermediate resonances. Therefore, an inclusion of intermediate resonances is necessary in order to quantitatively describe experimental angular distributions.

\subsubsection*{Comparison to lower energy}
A comparison of the normalized Legendre coefficients between this measurement and data collected at a lower value of excess energy $\epsilon$ = 162 MeV \cite{AbdelBary:2010pc} is listed \Cref{table:Legendre_compare}. The two sets of coefficients show striking differences for few coefficients indicating that the \Szero production mechanism changes between these values of excess energy. The CMS distributions are more forward-backward peaked for the proton and the \Szero hyperon and less peaked for the kaon, pointing to a larger relative contribution of pion with respect to kaon exchange at larger energies. In addition, the helicity angle distributions have a significant asymmetry at the highest energy, in contrast with the lower energy results.

\begin{table*}[t]
\centering
\caption{Comparison of the normalized Legendre coefficients between the present measurement and the data collected by COSY-TOF experiment at $\mathrm{\epsilon} = 162$ MeV \cite{AbdelBary:2010pc}.}
\label{table:Legendre_compare}
\begin{tabular}{*5c}
\toprule
 &  \multicolumn{2}{c}{$\epsilon$ = 162 MeV} & \multicolumn{2}{c}{$\epsilon$ = 556 MeV}\\
\midrule
{}   & $A_{1}/A_{0}$   & $A_{2}/A_{0}$     & $A_{1}/A_{0}$    & $A_{2}/A_{0}$ \\
$\mathrm{cos\theta^{CMS}_{\Sigma^{0}}}$   &  0.0 $\pm$ 0.0 & 0.03 $\pm$ 0.24  & 0.0 $\pm$ 0.0  & 0.32 $\pm$ 0.09\\
$\mathrm{cos\theta^{CMS}_{p}}$    &  0.0 $\pm$ 0.0 & 0.25 $\pm$ 0.29  & 0.0 $\pm$ 0.0 & 0.43 $\pm$ 0.13\\
$\mathrm{cos\theta^{CMS}_{K^{+}}}$   &  0.0 $\pm$ 0.0  &  0.48 $\pm$ 0.22   & 0.0 $\pm$ 0.0 & -0.01 $\pm$ 0.1\\
$\mathrm{cos\theta^{FR p\Sigma^{0}}_{p_{b,t}, p}}$   &  0.0 $\pm$ 0.0  &  0.11 $\pm$ 0.15   & -0.06 $\pm$ 0.17 & 0.27 $\pm$ 0.18\\
$\mathrm{cos\theta^{FR K^{+}\Sigma^{0}}_{p_{b,t}, K^{+}}}$   &  -0.04 $\pm$ 0.04  &  0.14 $\pm$ 0.18  & -0.19 $\pm$ 0.18  & 0.08 $\pm$ 0.19\\
$\mathrm{cos\theta^{FR K^{+}p}_{p_{b,t}, K^{+}}}$   &  -0.07 $\pm$ 0.07  &  0.57 $\pm$ 0.18   & 0.19 $\pm$ 0.12  & 0.05 $\pm$ 0.26\\
$\mathrm{cos\theta^{FR K^{+}\Sigma^{0}}_{p, \Sigma^{0}}}$   &  0.27 $\pm$ 0.27  &  -0.15 $\pm$ 0.15   & 0.36 $\pm$ 0.07  & -0.08 $\pm$ 0.09\\
$\mathrm{cos\theta^{FR p\Sigma^{0}}_{p, K^{+}}}$   &  -0.22 $\pm$ 0.22  &  0.0 $\pm$ 0.15  & -0.4 $\pm$ 0.06  & 0.04 $\pm$ 0.08\\
$\mathrm{cos\theta^{FR K^{+}p}_{K^{+}, \Sigma^{0}}}$  &  -0.11 $\pm$ 0.11  &  0.11 $\pm$ 0.18   & 0.55 $\pm$ 0.07  & -0.11 $\pm$ 0.09\\
\bottomrule
\end{tabular}
\end{table*}

\section{Total Cross Section}\label{xsection}
The total production cross section as function of the excess energy $\epsilon$ is used as a tool to compare the experimental data to the different theoretical approaches. The result on the \SSignal production cross section, obtained by integrating the $cos \mathrm{\theta^{*}_{\Sigma^{0}}}$ angular distribution, is

\begin{align*}
    \sigma (pK^{+}\Sigma^{0}) [\mu b] = 17.7 \pm 1.7 (stat) \pm 1.6 (syst) \:.
\end{align*}

\begin{figure}[!hbt]
    \centering
    \includegraphics[width=1.0\linewidth]{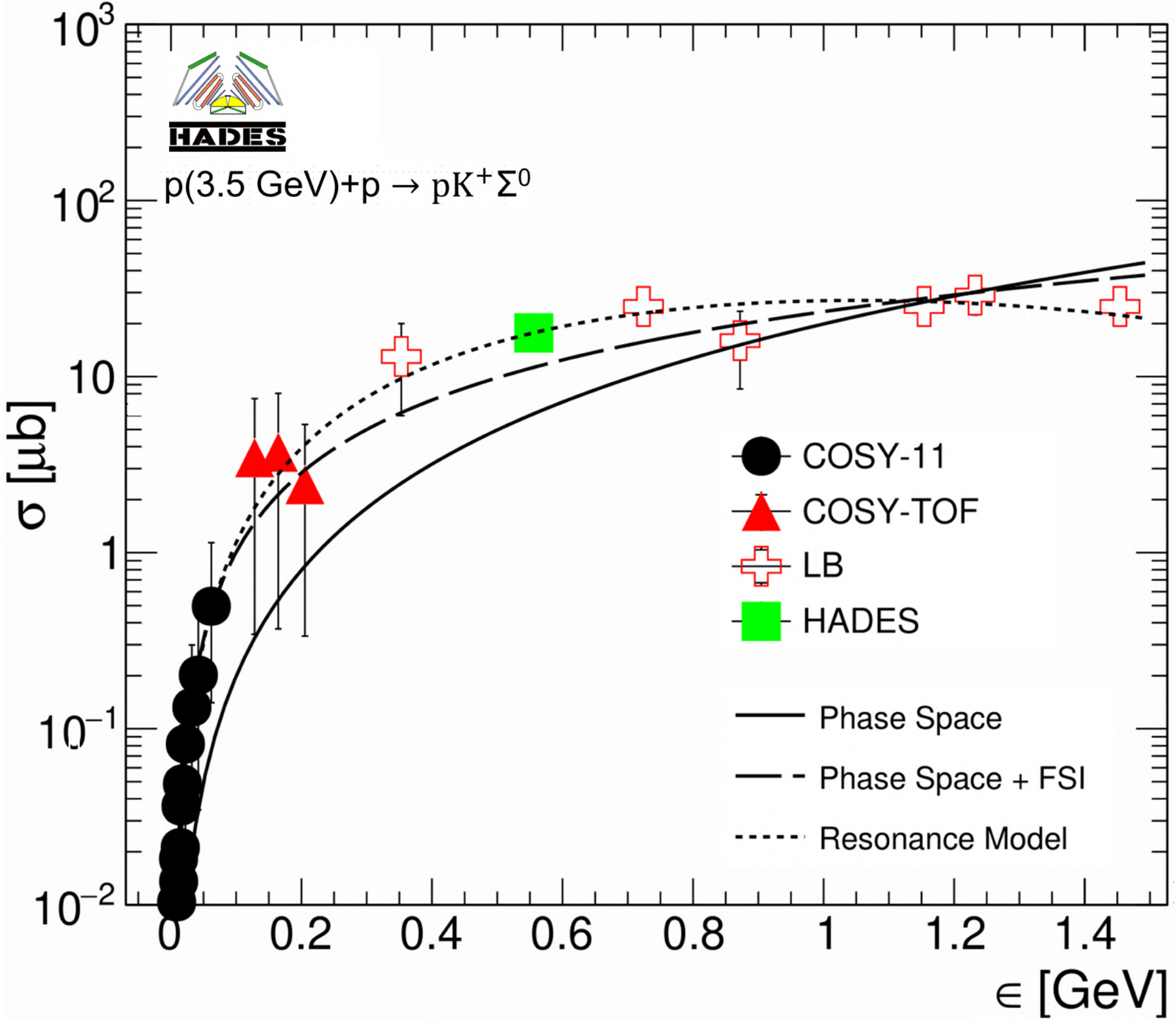}
    \caption{Compilation of cross sections of the reaction \SSignal from different experiments: COSY-11 \cite{GRZONKA199741, BALEWSKI199785, BALEWSKI1998211, Sewerin:1998ky, ABDELSAMAD200627,Valdau:2010wv}, COSY-TOF \cite{AbdelBary:2010pc} and data points from Landolt-B{\"o}rnstein (LB) \cite{Schopper:214138}. The production cross section of \Szero determined here is shown by the green square. The solid curve represents a pure phase space fit, the dotted curve is a parametrization based on the resonance model and the dashed curve is phase space and FSI as described in the text.}
    \label{figure:xsection}
    \end{figure}

The cross section value is included in \Cref{figure:xsection}, which shows a compilation of the \SSignal cross sections as a function of the excess energy. The present data point corresponds to $\epsilon$ = 556 MeV, which is depicted by the green square and existed in a region where no other measurements have been performed. This behaviour can not be described by phase space within experimental uncertainty as clearly seen by the solid curve $\sigma_{pK^{+}\Sigma^{0}} = K\epsilon^{2}$, where the quadratic excess-energy dependence is attributed to a pure (i.e. trivial) three-body phase space and $K$ is the fit free parameter.

An alternative parametrization proposed by F{\"a}ldt and Wilkin in \cite{Faldt:1996rh} that takes the proton-hyperon FSI interaction into account 

\begin{equation*}
    \sigma = C \cdot \frac{\epsilon^{2}}{(1+\sqrt{1+\epsilon/\alpha})^{2}} \:,
\end{equation*}

\noindent
where the parameters $C = 7.82 \times 10^{2} \mu b$ $GeV^{-2}$ and $\alpha = 4.57 \times 10^{-2} GeV$ are related to the FSI strength. Interestingly, the deviations to the pure phase space behavior start showing up at $\epsilon$ $>$ 200 MeV. The displayed data in that region could also be approximated by $\sigma$ $\approx$ 10 $\mu b$.

A more appropriate paramerization proposed by Tsushima in \cite{Tsushima:1998jz} shown by the dotted line is based on a resonance model, where the hyperon is produced via an intermediate nucleon resonance $\mathrm{N^{*}}$ or $\mathrm{\Delta^{*}}$. This paramerization describes all data points near threshold up to 1.4 GeV fairly well.

\begin{figure}[!hbt]
    \centering
    \includegraphics[width=1.0\linewidth]{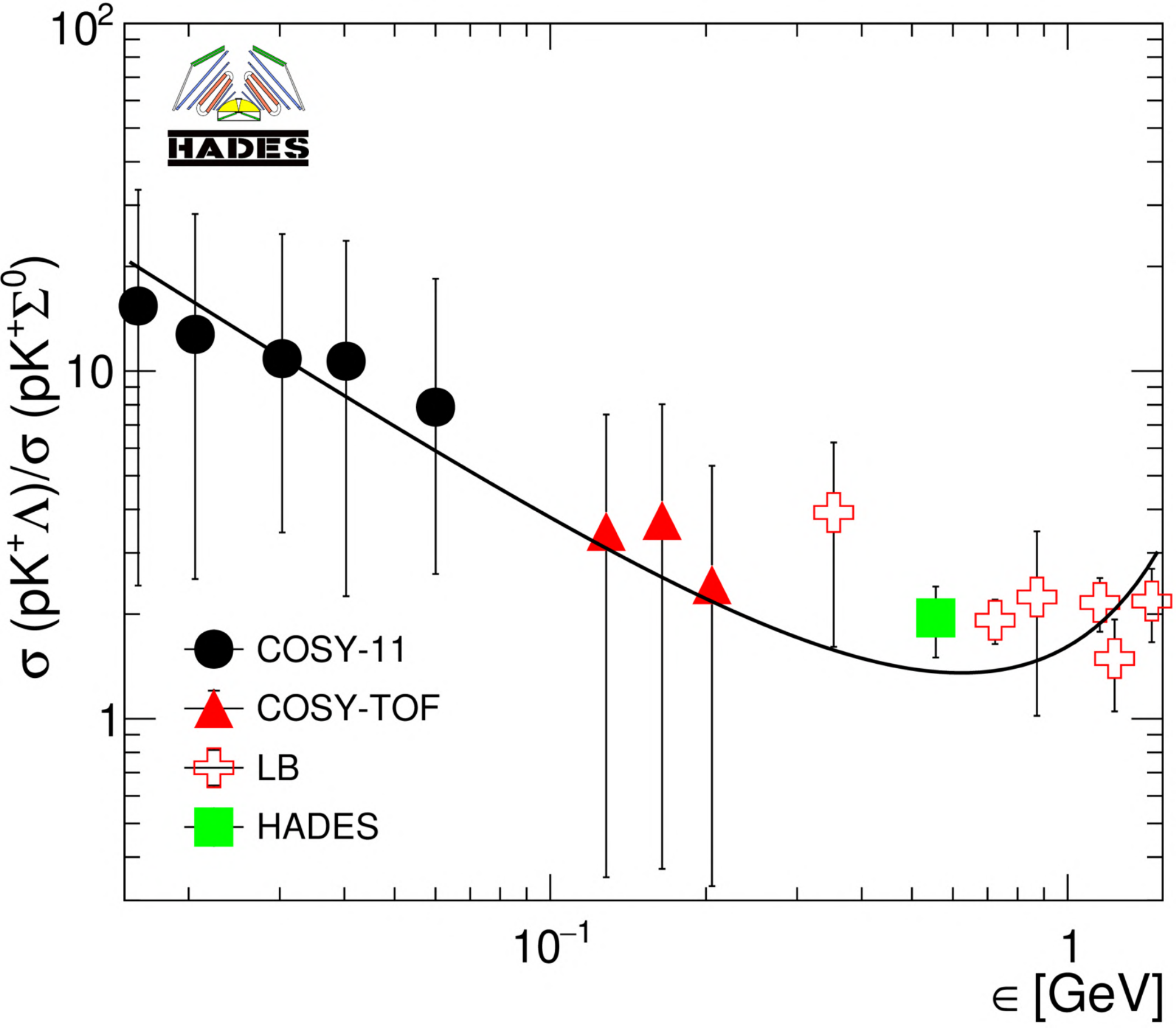}
    \caption{Experimental cross section ratio of the present data point together with a compilation of the world data: COSY-11 \cite{GRZONKA199741, BALEWSKI199785, BALEWSKI1998211, Sewerin:1998ky, ABDELSAMAD200627,Valdau:2010wv}, COSY-TOF \cite{AbdelBary:2010pc} and data points from Landolt-B{\"o}rnstein (LB) \cite{Schopper:214138}. The present data square is shown by the green square. The solid curve is the ratio of the paramerization of both channels \cite{Faldt:1996rh, Tsushima:1998jz}.}
    \label{figure:xsection_ratio}
    \end{figure}

Using the $\mathrm{pp \rightarrow pK^{+}\Lambda}$ cross section measured by the HADES collaboration \cite{Agakishiev:2014dha}, the cross section ratio $\sigma(pK^{+} \Lambda)/\sigma(pK^{+}\Sigma^{0})$ is determined to be 1.90 $\pm$ 0.41. Based on the coupled channel calculation, where the interference of the pion and kaon exchange is taken in account, the cross section ratio can be reproduced by selecting the relative sign for these two mechanism \cite{Kowina:2004kr}. \Cref{figure:xsection_ratio} shows the cross section ratio as a function of the excess energy together with a compilation of other measurements \cite{Schopper:214138}. The solid curve is the ratio of the paramerization of both channels, where the paramerization proposed by F{\"a}ldt and Wilkin \cite{Faldt:1996rh} based on phase space and FSI is used for \LSignal and the Tsushima paramerization \cite{Tsushima:1998jz} based on a resonance model is used for the \SSignal channel.

The observed cross section ratio in the present p+p data is similar to the corresponding value measured in p+Nb data \cite{Adamczewski-Musch:2017gjr}, despite the large difference in the individual cross sections, thus corroborating the importance of FSI for these reactions.

\section{Partial Wave Analysis}\label{section:pwa}
From the results presented above, it was concluded that the experimental data on angular distributions can not be described by pure phase space production, but there must be a resonant component as anticipated in \cite{AbdelBary:2010pc}. Therefore, a Partial Wave Analysis (PWA) using the Bonn-Gatchina Partial Wave Analysis (Bo-Ga PWA) framework \cite{Sarantsev:2005tg} has been applied with the goal to quantify the relative contributions of different partial waves.

The Bo-Ga PWA framework takes a list of possible transition waves as an input that may contribute to the final state. The non-resonant production proceeds as follows: the proton (\spin = $\frac{1}{2}^{+}$) and the hyperon (in this case \Szero with \spin = $\frac{1}{2}^{+}$) are combined into a two particle sub-system and then the kaon (\spin = $0^{-}$) is combined with this sub-system to produce the three-body final state. In case of the resonant production, the proton is combined with one of the resonances listed in \Cref{table:list_resonances} $\mathrm{N^{*}}$-p, or $\mathrm{\Delta^{*}}$-p to produce the final state \SSignal. Resonance masses and widths were fixed to the PDG values \cite{Tanabashi:2018oca} in order to reduce the number of the free fit parameters.

\begin{table}[t]
 \caption{A list of $\mathrm{N^{*}}$ and $\mathrm{\Delta^{*}}$ resonances that might contribute to the \SSignal reaction. The mass, width and spin-parity quantum numbers were taken from \cite{Tanabashi:2018oca}.}
 \label{table:list_resonances}
 \begin{tabularx}{1.0\linewidth}{L{1.0}C{0.6}C{0.6}C{0.6}}
 \toprule
  Resonance & Mass [\GeV] & Width [\GeV] & \spin \\
 \midrule
     $\mathrm{N^{*}(1710)}$ & 1.710 & 0.140 & $\frac{1}{2}^{+}$ \\ 

     $\mathrm{N^{*}(1875)}$ & 1.875 & 0.200 & $\frac{3}{2}^{-}$ \\ 

     $\mathrm{N^{*}(1880)}$ & 1.880 & 0.300 & $\frac{1}{2}^{+}$  \\

    $\mathrm{N^{*}(1895)}$ & 1.895 & 0.120 & $\frac{1}{2}^{-}$ \\

     $\mathrm{N^{*}(1900)}$ & 1.920 & 0.200 & $\frac{3}{2}^{+}$ \\

     $\mathrm{\Delta^{*}(1900)}$ & 1.860 & 0.250 & $\frac{1}{2}^{-}$ \\
 
    $\mathrm{\Delta^{*}(1910)}$ & 1.900 & 0.300 & $\frac{1}{2}^{+}$ \\
      
     $\mathrm{\Delta^{*}(1920)}$ & 1.920 & 0.300 & $\frac{3}{2}^{+}$ \\
 \bottomrule
 \end{tabularx}
\end{table}

The strength ($\mathrm{\alpha_{1}}$) and the phase ($\mathrm{\alpha_{2}}$) of each transition wave are determined by fitting the partial wave amplitudes to the experimental data on an event-by-event basis in an unbinned fit. The fit is based on a log-likelihood minimization and the fitting procedure is repeated for many iterations until there is no further improvement of the log-likelihood value. By comparing the log-likelihood value of many fits the best fit can be determined through the largest negative value. As an output, the BG-PWA returns the fitted values of the parameters $\mathrm{\alpha_{1}}$ and $\mathrm{\alpha_{2}}$ and a list of simulated events that have been used as an input but with each event being assigned a weight factor, which gives the contribution of this event to the total yield.

Since the signal region contains background events (mainly \LSignal and \LBack), and because the Bo-Ga PWA method works on an event-by-event basis, it is important to identify whether a particular event belongs to the signal or the background. The \LSignal contribution is three times larger than \LBack inside the signal region. Therefore, the \LSignal channel is considered the main contributing background and its kinematics is modeled by performing a PWA on the \LSignal-like events. The solutions published in \cite{Agakishiev:2014dha} have been tested and solution No. 8/1 was found to provide the best description of the experimental data by including the p+p initial waves ${}^{2S+1}L_{J}$ = ${}^{1}S_{0}$, ${}^{3}P_{0}$, ${}^{3}P_{1}$ and ${}^{1}D_{2}$.

The solution No. 8/1 is then applied to the \Lzero 4$\pi$-phase space simulations and these events are filtered through the full simulation and analysis chain. After reconstructing the \Lzero events that have been assigned a PWA weight, the missing mass $\mathrm{MM(pK^{+})}$ spectrum was investigated and the \Lzero contribution in the signal region $\mathrm{1.170 < MM(pK^{+}) [GeV/c^{2}]< 1.220}$ was determined to be 292 events. Those events are then added to the signal list with a negative weight.

After subtracting the \Lzero contribution, the PWA technique is applied to the \SSignal events. A systematic variation of the input partial waves was performed and, in addition, the number of non-resonant and resonant final partial waves was varied and the quality of the PWA solution was determined by the negative log-likelihood value of the fit.

The best PWA solution shown by the dashed histograms in Figures \ref{figure:angularD} and \ref{figure:Yang} was obtained by including p+p initial waves ${}^{2S+1}L_{J}$ = ${}^{2}S_{0}$, ${}^{3}P_{0}$, ${}^{3}P_{1}$, ${}^{3}P_{2}$, ${}^{1}D_{2}$ and ${}^{3}F_{2}$. In addition, nucleon resonances $\mathrm{N^{*}(1710)}$, $\mathrm{N^{*}(1900)}$ and $\mathrm{\Delta^{*}(1900)}$ were found to contribute as well as non-resonant partial waves. However, due to the limited statistics and the large number of free fit parameters, an unambiguous determination of the contributions of each resonance is not possible since these contributions vary significantly for different solutions. Nevertheless, resonances with masses around 1.710 \GeV ($\mathrm{N^{*}(1710)}$) and 1.900 \GeV ($\mathrm{N^{*}(1900)}$ or $\mathrm{\Delta^{*}(1900)}$) are certainly preferred by the fit. 

\section{Conclusion and Outlook}\label{sec:summary}
The exclusive reconstruction of the reaction \SSignal at a beam kinetic energy of 3.5 GeV has been presented and the \SSignal total production cross section was determined with an accuracy better than 10 \% in a region where no data existed. The dynamics of the reaction was investigated by studying the angular distributions in the CMS, G-J and helicity frame. The corrected CMS distributions of the hyperon and the proton show anisotropies, which it is more pronounced in the case of the proton. This is the expected behavior if the pion exchange mechanism dominates the particle production process in a simple one-boson exchange formalism. In addition, an investigation of the \Szero T-Y angle measured in the $\mathrm{K^{+}\Sigma^{0}}$ reference frame, deviates from isotropy, which hints to a non-negligible contribution of the of kaon exchange mechanism.

The helicity angular distributions are not isotropic, which indicates that a pure phase space description without momentum-dependent matrix element(s) is by far not appropriate. The influence of different nucleon resonances has been tested by means of a PWA using the Bo-Ga PWA framework. The best solution was obtained by including the initial p+p configuration ${}^{1}S_{0}$, ${}^{3}P_{0}$, ${}^{3}P_{1}$, ${}^{3}P_{2}$, ${}^{1}D_{2}$ and ${}^{3}F_{2}$. Due to the limited statistics, it was not possible to obtain the exact strength of the individual nucleon resonances. However, nucleon resonances $\mathrm{N^{*}(1710)}$, $\mathrm{N^{*}(1900)}$ and $\mathrm{\Delta^{*}(1900)}$ are preferred by the fit.

Recently, the HADES setup has been upgraded by an electromagnetic calorimeter (ECAL) and a Forward Detector (FD) based on PANDA experiment straw tubes \cite{Adamczewski-Musch:2021rlv}. The new data that was collected in February 2022 offers the opportunity to perform the same measurement with an upgraded setup at a higher proton beam energy of 4.5 GeV. This upgrade will allow the identification of the daughter photon in $\mathrm{\Sigma^{0} \rightarrow \Lambda \gamma}$ via the ECAL. In addition, it will improve the mass resolution of the \Lzero hyperon in the FD acceptance and consequently improve the quality of the kinematic refit. Furthermore, the collected data will provide sufficient statistics to extract quantitative contributions of the different nucleon resonances and a measurement of their $\mathrm{K^{+}\Sigma^{0}}$ branching ratios, which will certainly improve the current measurement.
 
\begin{acknowledgement}
\section{Acknowledgment}
The HADES collaboration gratefully acknowledges the support by SIP JUC Cracow, Cracow (Poland), 2017/26/M/ST2/00600; WUT Warsaw (Poland) No: 2020/38/E/ST2/00019 (NCN), IDUB-POB-FWEiTE-3; TU Darmstadt, Darmstadt (Germany), VH-NG-823, DFG GRK 2128, DFG CRC-TR 211, BMBF:05P18RDFC1, HFHF, ELEMENTS 500/10.006, GSI F\&E, EMMI at GSI Darmstadt; Goethe-University, Frankfurt (Germany), BMBF:05P12RFGHJ, GSI F\&E, HIC for FAIR (LOEWE), EMMI at GSI Darmstadt; JLU Giessen, Giessen (Germany),BMBF:05P12RGGHM; IJCLab Orsay, Orsay (France), CNRS/IN2P3; NPI CAS, Rez, Rez (Czech Republic), MSMT LTT17003, MSMT LM2018112, MSMT OP VVV CZ.02.1.01/0.0/0.0/18\_046/0016066; European Union’s Horizon 2020, no. 824093 (STRONG2020). \\
This project has received funding from the programme "Netzwerke 2021", an initiative of the Ministry of Culture and Science of the State of Northrhine Westphalia. The sole responsibility for the content of this publication lies with the authors. \\
The following colleagues from Russian institutes did contribute to the results presented in this publication but are not listed as authors following the decision of the HADES Collaboration Board on March 23, 2022: G. Agakishiev, A. Belyaev, O. Fateev, A. Ierusalimov, V. Ladygin, T. Vasiliev, M. Golubeva, F. Guber, A. Ivashkin, T. Karavicheva, A. Kurepin, A. Reshetin, A. Sadovsky and A.V.Sarantsev.

\end{acknowledgement}

\printbibliography

\end{document}